\documentclass[
 aip,
 jcp,
 amsmath,amssymb,
 reprint,
citeautoscript,
]{revtex4-1}

\usepackage{graphicx}
\usepackage{dcolumn}
\usepackage{bm}
\usepackage{braket}
\usepackage{mathtools}

\DeclarePairedDelimiter\abs{\lvert}{\rvert}

\begin{document}

\title{Efficient construction of generalized master equation memory kernels for multi-state systems from nonadiabatic quantum-classical dynamics}

\author{William C. Pfalzgraff}
\affiliation{Department of Chemistry, Stanford University, Stanford, California, USA}
\affiliation{Department of Chemistry, Chatham University, Pittsburgh, Pennsylvania, USA}
\author{Andr\'{e}s Montoya-Castillo}
\affiliation{Department of Chemistry, Stanford University, Stanford, California, USA}
\author{Aaron Kelly}
\affiliation{Department of Chemistry, Dalhousie University, Halifax, Nova Scotia, Canada}
\author{Thomas E. Markland} 
   \email{tmarkland@stanford.edu}
\affiliation{Department of Chemistry, Stanford University, Stanford, California, USA}
\date{\today}

\begin{abstract}
Methods derived from the generalized quantum master equation (GQME) framework have provided the basis for elucidating energy and charge transfer in systems ranging from molecular solids to photosynthetic complexes. Recently, the non-perturbative combination of the GQME with quantum-classical methods has resulted in approaches whose accuracy and efficiency exceed those of the original quantum-classical schemes while offering significant accuracy improvements over perturbative expansions of the GQME. Here we show that, while the non-Markovian memory kernel required to propagate the GQME scales quartically with the number of subsystem states, the number of trajectories required scales at most quadratically when using quantum-classical methods to construct the kernel. We then present an algorithm that allows further acceleration of the quantum-classical GQME by providing a way to selectively sample the kernel matrix elements that are most important to the process of interest. We demonstrate the utility of these advances by applying the combination of Ehrenfest mean field theory with the GQME (MF-GQME) to models of the Fenna-Matthews-Olson (FMO) complex and the light harvesting complex II (LHCII), with 7 and 14 states, respectively. This allows us to show that MF-GQME is able to accurately capture all the relevant dynamical time-scales in LHCII: the initial nonequilibrium population transfer on the femtosecond time-scale, the steady state-type trapping on the picosecond time-scale, and the long time population relaxation. Remarkably, all of these physical effects spanning tens of picoseconds can be encoded in a memory kernel that decays after only $\sim$65 fs.
\end{abstract}


\maketitle

\section{Introduction}
Theoretical modelling of processes ranging from energy transfer in photosynthetic light harvesting complexes and organic electronics to proton coupled electron transfer, require the development of methods that allow for the accurate and efficient calculation of the dynamics of open quantum systems, i.e., a subsystem of interest coupled to an external environment. Generalized quantum master equations (GQMEs) provide a formally exact way of rewriting the equation of motion of a set of degrees of freedom of particular interest via the use of the appropriate projection operator \cite{Nakajima1958,Zwanzig1960a,berne_book}. For example, in open quantum systems, the projection operator is frequently chosen to encompass all the states of the electronic subsystem, although this choice need not align with simple partitions at the level of the Hamiltonian, i.e., the projector could contain only a subset of electronic states or even include some nuclear degrees of freedom. The effect of the remaining degrees of freedom on the dynamics of those of interest is encoded through the memory kernel.\cite{Nakajima1958, Zwanzig1960a, Mori1965} However, evaluating the memory kernel requires overcoming the complications that arise from the presence of the projected propagator. 

A common approach to avoid the projected propagator is to treat the memory kernel using second order perturbation theory. This leads to methods including Markovian and non-Markovian Redfield theory and its variants \cite{Bloch1957, Redfield1965, Chang1993, mukamel_redfield}, the noninteracting blip approximation (NIBA) \cite{Leggett1987, Dekker1987}, F\"orster theory \cite{Forster1960}, and the polaronic quantum master equation \cite{Jang2008,Jang2011a, Nazir2009, McCutcheon2011}, which have been used to treat a wide variety of multi-state electronic energy transfer systems. In these approaches, the memory kernel depends only on correlation functions of the isolated system and bath, and hence can be straightforwardly generated even for systems with many subsystem states.\cite{BreuerPetruccione,NitzanBook} However, these approaches are limited in their applicability to systems where the perturbative parameter remains small.

More recently, it has been shown that the projected propagator can be exactly removed by employing the Dyson identity, leading to a representation of the full (non-perturbative, non-Markovian) memory kernel in terms of \emph{unprojected} correlation functions of the system and bath.\cite{Shi2003} This allows one to gain accurate results over a much wider range of regimes than perturbative master equations, albeit at the expense of requiring cross-correlation functions of system and bath operators to generate the memory kernel. Evaluating these cross-correlation functions using trajectory-based quantum-classical approaches with the GQME formalism has been shown to give considerable increases in both accuracy and efficiency compared to using the same quantum-classical method alone.\cite{Shi2004a,Kelly2013, Kelly2015, Pfalzgraff2015, Montoya2016a, KellyMontoya2016, Montoya2017b}

However, while these studies showed the efficacy of such a combination in systems containing two subsystem states, for a system with $N_{s}$ subsystem states, the reduced density matrix (RDM) has $N_{s}^2$ elements and hence the size of the memory kernel needed to propagate it grows as $N_{s}^4$. One might therefore imagine that combining quantum-classical methods with the GQME would be limited to problems with small numbers of subsystem states. Here, we demonstrate that for quantum-classical methods that treat the subsystem using wavefunctions, e.g., Ehrenfest mean field theory (MFT) \cite{mft,mft2,mft3} and fewest switches surface hopping (FSSH) \cite{tully1,tully2,Hammes-Schiffer1996}, that the full memory kernel can be obtained using only $\mathcal{O}(N_{s}^2)$ trajectories. 

We then show how, in cases where one is interested in a particular (or a small set of) subsystem initial conditions, one can further improve the efficiency by focusing resources into accurately capturing the matrix elements of the memory kernel that most strongly determine the evolution of a given initial condition. We achieve this by noting that, while the fully converged memory kernel contains the information necessary propagate the GQME from any subsystem initial condition contained in the projector, often many of these may not be of interest. Here we introduce an algorithm that exploits this observation to identify and selectively converge the kernel elements that are most important to capture the relaxation from a given subsystem initial condition. This allows one to minimize the computational effort expended on the elements that only exert a minor influence on the process of interest. 

We then illustrate the scalability of the quantum-classical GQME to multi-state systems by applying the mean field GQME (MF-GQME) method~\cite{Kelly2015} to treat the 7-state Fenna-Matthews-Olsen (FMO) complex~\cite{FMO1,Ishizaki2009b} and the 14-state light harvesting complex II (LHCII).\cite{Novoderezhkin2005,lhcii_model2,lhcii_model3} We show that MF-GQME gives quantitative accuracy in regimes where direct MFT and modified Redfield theory fail, while remaining less expensive than a direct application of MFT even in these larger systems. This development thus opens the door to the application of quantum-classical GQME approaches to multi-state systems coupled to atomistic environments that lie beyond linear system-bath coupling and harmonic environments.\cite{Pfalzgraff2015} 

\section{Theory}
\label{sec:theory}
To show how the GQME formalism can be efficiently used with trajectory-based quantum dynamics methods in multi-state systems, here we give the general expression for the memory kernel for a system with arbitrary system-bath coupling. By analyzing the form of this general expression, we demonstrate that only $\mathcal{O}(N_s^2)$ trajectories are required to calculate the memory kernel in the MF-GQME method in a system with $N_s$ subsystem states. 

To begin, we consider a Hamiltonian of the general form,
\begin{equation}\label{eq:general-open-quantum-system-hamiltonian}
    \hat{H} = \hat{H}_s + \hat{H}_b + \hat{H}_{sb},
\end{equation}
where $\hat{H}_s$ is the isolated subsystem Hamiltonian, $\hat{H}_b$ the isolated bath Hamiltonian, and $\hat{H}_{sb}$ contains all the terms that couple the system and the bath. For any $\hat{H}_{sb}$, one can write it such that it is in the form of a sum of direct products of system and bath operators,
\begin{equation}
    \hat{H}_{sb} = \sum_{j=1}^{N_{sb}} \hat{S}_j \otimes \hat{\Gamma}_j,
    \label{eq:general_H_sb}
\end{equation}
where $\hat{S}_j$ is a pure subsystem operator, $\hat{\Gamma}_j$ is a pure bath operator, and $N_{sb}$ is the number of terms in the sum. 

Here we consider initial conditions for the full system density operator $\hat{\rho}(0)$ that are of the spectroscopic form, $\hat{\rho}(0) = \hat{\rho}^{eq}_b \otimes \hat{\rho}_s(0)$, where $\hat{\rho}_s(0)$ encodes the initial state of the subsystem and $\hat{\rho_b}^{eq}$ is the equilibrium bath density operator, $\hat{\rho}^{eq}_b = e^{-\beta \hat{H}_b} / \mathrm{Tr}_b \left\{ e^{-\beta \hat{H}_b}\right\}$. While this choice of initial condition simplifies the expressions for the memory kernel, we emphasize that the scaling arguments presented here do not depend on this choice and are straightforward to generalize to any factorizable initial condition.

The quantity of interest is the RDM, $\hat{\rho}_s(t) = \mathrm{Tr}_b\{ \hat{\rho} (t) \}$, from which any subsystem observable, such as the population relaxation or electronic spectra, can be generated. In a given basis, one can write the RDM as
\begin{equation}
    \hat{\rho}_s(t) = \sum_{n=1}^{N_s^2} [\rho_s]_n (t) \hat{A}_n,
\end{equation}
where $\hat{A}_n \in \left\{ \ket{\alpha}\bra{\alpha^{\prime}} \right\}$, which is a set of $N_s^2$ subsystem operators that span the subsystem Liouville space,
 and $[\rho_s]_n$ denotes a particular matrix element of the RDM. These matrix elements can be expressed as quantum mechanical expectation values,
  \begin{equation}
     \left[ \hat{\rho}_s(t) \right]_n = \mathrm{Tr} \left\{ \hat{\rho}(0) e^{i\mathcal{L}t/\hbar}\hat{A}_n \right\} = \mathrm{Tr}\left\{\hat{\rho}(t) \hat{A}_n \right\}.
 \end{equation}
In the Mori-Nakajima-Zwanzig formalism, the dynamics of the RDM can be obtained by integrating the GQME \cite{Zwanzig1960a,Nakajima1958,Mori1965,berne_book,BreuerPetruccione},
\begin{equation}
    \dot{ \rho}_s(t)=\rho_s(t)\mathcal{X} -\int^t_0 d\tau \ \rho_s(t-\tau)\mathcal{K}(\tau) ,
\label{eq:rdm_evol}
\end{equation}
 where $\rho_s(t)$ is a row vector with $N_s^2$ time-dependent elements and $\mathcal{X}$ and $\mathcal{K}(t)$ are matrices of size $N_s^2 \times N_s^2$, and  $\dot{ \rho}_s(t) = \frac{d}{dt} \rho_s(t)$. The static matrix $\mathcal{X}_{mn}$ = $\frac{i}{\hbar} \mathrm{Tr_s}\left\{ \hat{A}_m^{\dagger}\mathcal{L}_s \hat{A}_n \right\}$ ensures that the GQME exactly recovers the free subsystem evolution when the memory kernel, $\mathcal{K}(t)$, is zero, i.e.\ when the subsystem and bath do not interact. The memory kernel, which encodes the effect of the bath on the subsystem dynamics, can be constructed using correlation functions in the full system and bath Hilbert space. Using the Dyson decomposition of the propagator\cite{evans_moriss} one obtains,\cite{Shi2003,Montoya2016a} 
\begin{equation}
    \mathcal{K}(t) = \mathcal{K}_1(t) - i\int_0^t d\tau \ \mathcal{K}_3(t-\tau)\mathcal{K}(\tau).
    \label{eq:K_K1_K3}
\end{equation}
Here $\mathcal{K}_3$ and $\mathcal{K}_1$ are auxiliary kernels, which can be obtained using one's choice of dynamical method, and the subscripts 1 and 3 are used to be consistent with earlier work.\cite{Shi2003,Shi2004a,Kelly2013,Kelly2015,Pfalzgraff2015,Montoya2016a} For any system-bath coupling, $\mathcal{K}_3$ and $\mathcal{K}_1$ can be written as linear combinations of correlation functions of system and bath operators of the form (see Appendix \ref{app:auxiliary-kernels}),
\begin{align}
        \mathcal{K}_3(t)  &= \frac{1}{\hbar}\sum_{j=1}^{N_{sb}} \Big[ c^{(j,-)} q_{3}^{(j,+)}(t)  + c^{(j,+)}  q_{3}^{(j,-)}(t)\Big],
        \label{eq:general_K3}\\
         \mathcal{K}_1(t)  &= \frac{1}{\hbar^2}\sum_{j,k=1}^{N_{sb}}  \Big[ c^{(j,-)}  q_{1}^{(jk,+)}(t)c^{(k,-)} \nonumber   \\ 
        &\qquad \qquad \qquad \quad + c^{(j,+)}  q_{1}^{(jk,-)}(t)c^{(k,-)} \Big].
         \label{eq:general_K1}
\end{align}
Here, like the memory kernels, $c^{(j,\pm)}$, $q_{3}^{(j,\pm)}(t)$, and $q_{1}^{(jk,\pm)}(t)$ are matrices of size $N_s^2 \times N_s^2$. The transformation matrices $c^{(j,\pm)}$ are time-independent and simple to evaluate analytically, with their elements given by,
\begin{equation}
    c_{mn}^{(j,\pm)} = \mathrm{Tr_s} \left\{ \hat{A}_m^{\dagger} \left[\hat{S}_j,\hat{A}_n\right]_{\pm} \right\},
    \label{eq:transformation_elements}
\end{equation}
where $[\cdot, \cdot]_{-}$ denotes the commutator and $[\cdot, \cdot]_{+}$ the anti-commutator. The matrix elements of the correlation functions $q_{3}^{(j,\pm)}(t)$ and $q_{1}^{(jk,\pm)}(t)$ are (see Appendix \ref{app:auxiliary-kernels})
\begin{align}
    \label{eq:general_q3}
    q_{3,mn}^{(j,\pm)}(t) &= \frac{1}{2}\mathrm{Tr}\left\{ \left[ \hat{\rho}_{b}^{eq}, \hat{\Gamma}_j\right]_{\pm} \hat{A}_m^{\dagger} e^{i\mathcal{L}t/\hbar} \hat{A}_n  \right\} ,\\
    \label{eq:general_q1}
    q_{1,mn}^{(jk,\pm)}(t) &= \frac{1}{2}\mathrm{Tr}\left\{ \left[ \hat{\rho}_{b}^{eq}, \hat{\Gamma}_j\right]_{\pm} \hat{A}_m^{\dagger} e^{i\mathcal{L}t/\hbar} \hat{\Gamma}_{k} \hat{A}_n  \right\}. 
\end{align}
Equations (\ref{eq:general_K3}) and (\ref{eq:general_K1}) provide a formal basis for the calculation of the memory kernel, in which one obtains the auxiliary kernels $\mathcal{K}_3$(t) and $\mathcal{K}_1$(t) by evaluating $c^{(j,\pm)}$ exactly and calculating $q_{3}^{(j,\pm)}(t)$ and $q_{1}^{(jk,\pm)}(t)$ using one's choice of exact or approximate dynamical method. These quantities are then used to to obtain the auxiliary kernels $\mathcal{K}_1$ and $\mathcal{K}_3$ using Eqs.(\ref{eq:general_K3})--(\ref{eq:general_K1}), and the full memory kernel is obtained using Eq.\ (\ref{eq:K_K1_K3}) (see Appendix \ref{app:K_K1_K3}) In Sec.\ \ref{ssec:scaling}, we use these expressions to analyze the scaling with the number of subsystem states.

\subsection{Scaling of the MF-GQME method with number of subsystem states}
\label{ssec:scaling}

In the MF-GQME method, one approximates the correlation functions defined in Eqs.\ (\ref{eq:general_q3}) and (\ref{eq:general_q1}) using MFT (see Appendix \ref{app:mft_dynamical_matrices}). The memory kernel can then be used to solve the GQME for the dynamics of the RDM, given by Eq.\ (\ref{eq:rdm_evol}). In practice, the computational cost of generating the full memory kernel using Eqs.\ (\ref{eq:K_K1_K3})--(\ref{eq:general_K1}) and integrating the GQME is negligible compared to the cost of generating the MFT trajectories necessary to calculate $\mathcal{K}_3$ and $\mathcal{K}_1$. The computational cost is therefore determined by the number of MFT trajectories required to evaluate $q_{3}^{(j,\pm)}(t)$ and $q_{1}^{(jk,\pm)}(t)$, and how this cost scales with the number of subsystem states. 

The number of correlation functions that need to be calculated to evaluate $q_{3}^{(j,\pm)}(t)$ and $q_{1}^{(jk,\pm)}(t)$ using Eqs.\ (\ref{eq:general_q3}) and (\ref{eq:general_q1}) can at first appear formidable, as the indices $m$ and $n$ in those equations each run over $N_s^2$ elements. This might lead one to erroneously conclude that the MF-GQME method scales as $\mathcal{O}(N_s^4)$, which would severely limit its applicability to treat problems with many subsystem states. However, closer inspection of these expressions in the context of MFT dynamics reveals that one only needs to sample the distinct electronic initial conditions, labelled by the index $m$, to calculate \emph{all} the necessary correlation functions. This is because the MFT trajectories generated from a given electronic initial condition, $\hat{A}_m^{\dagger}$, provide the time-dependent observables necessary to calculate all of the correlation functions corresponding to that initial condition. Consequently, since only $N_s^2$ electronic initial conditions need to be sampled, only  $\mathcal{O}(N_s^2)$ quantum-classical trajectories are required. In practice, for MFT the number of initial conditions can be shown to be $\frac{1}{2} N_s(N_s+1)$ by following the procedure outlined in Appendix \ref{App:off_diagonal_inicon}, which exploits the fact that the correlation functions corresponding to a given off-diagonal initial condition and its Hermitian conjugate can be obtained from the same set of trajectories. 

In general, evaluating forces dominates the overall cost, especially in atomistic systems or systems with a large number of bath degrees of freedom. However, the calculation of the correlation functions required scales as $\mathcal{O}(N_s^4)$ and integration of the GQME scales as $\mathcal{O}(N_s^6)$ which could dominate the cost for large $N_s$. However, even for the systems studied here which use very simple forms of the bath, the cost of integrating the GQME is negligible.

An additional concern regarding the scalability of the MF-GQME approach comes from amount of memory required to store all elements in $q_{3,mn}^{(j,\pm)}(t)$ and $q_{1,mn}^{(j,\pm)}(t)$. For instance, the number of time-dependent elements in $q_{1,mn}^{(j,\pm)}(t)$ can be calculated as the product of $N_s^2$ initial electronic conditions $\hat{A}_m^{\dagger}$, $N_{sb}$ bath operators $\hat{\Gamma}_j$ measured at $t=0$, and $N_s^2$ electronic operators $\hat{A}_n(t)$ and $N_{sb}$ bath operators $\hat{\Gamma}_j(t)$ measured at finite times. This corresponds to $\mathcal{O}(N_{s}^4 \times N_{sb}^2)$ time dependent elements that need to be stored in memory, and hence one might expect the amount of available memory to be a significant limitation. In practice, however, this problem can be easily circumvented because $q_{3}^{(j,\pm)}(t)$ and $q_{1}^{(jk,\pm)}(t)$ do not themselves need to be stored in memory. Instead, one can evaluate their contributions to $\mathcal{K}_3(t)$ and $\mathcal{K}_1(t)$ on the fly using Eqs.\ (\ref{eq:general_K3}) and (\ref{eq:general_K1}), so that only $\mathcal{K}_3(t)$ and $\mathcal{K}_1(t)$ need to be stored. In addition, because not all of the matrix elements are independent, there are only $\frac{1}{2}N_s^2(N_s^2+1)$ unique elements in each,  with the remainder being complex conjugates of these elements.\cite{Shi2003} The amount of memory required to store the kernels is therefore proportional to $N_s^2(N_s^2+1) \times N_{steps}$ where $N_{steps}$ is the number of timesteps before the kernel decays. In practice, when the memory kernel decays quickly to zero $N_{steps}$ is a small number,  meaning that negligible storage is required.

\subsection{Accelerating convergence of the GQME using selective sampling of kernel matrix elements}
\label{ssec:selective}

Once the memory kernel has been calculated, solving the GQME, Eq.\ (\ref{eq:rdm_evol}), provides access to the subsystem dynamics for any factorizable initial condition of the subsystem RDM. However, in many cases one is only interested in a particular initial condition, or a set of them. For example, one may be interested in the relaxation dynamics resulting from the initial excitation of a particular chromophore in a molecular system consisting of multiple chromophores. Here it is important to make a distinction between the initial condition of interest at the RDM level, and the electronic initial conditions that need to be sampled in the MFT calculation of the memory kernel. At the memory kernel level, all separable initial conditions (e.g., initial excitation of each chromophore and coherence between chromophores) need to be sampled to construct the full memory kernel. It is this complete sampling of the initial conditions at the memory kernel level that ensures that one can use Eq.\ (\ref{eq:rdm_evol}) to propagate the RDM dynamics from \emph{any} separable initial condition. 

However, as in the case mentioned above, suppose that one is only interested in propagating the RDM starting from one particular initial condition (or a small subset of initial conditions) in a system with a large number of subsystem states. In this case, there may be many pathways that do not play a role in the relaxation dynamics from the RDM initial condition of interest. Hence, it is clearly unnecessary to have accurate knowledge of the memory kernel elements that describe these unimportant relaxation pathways.

One can exploit this realization by focusing resources into converging the elements of the memory kernel that are most important to describe the relaxation process of interest. When the memory kernel elements are generated using MFT, this corresponds to identifying and selectively converging the most important correlation functions from the sets $\{q_{3}^{(j,\pm)}(t)\}$ and $\{q_{1}^{(jk,\pm)}(t)\}$  (defined in Eqs.~(\ref{eq:general_q3}) and (\ref{eq:general_q1})), which give rise to the memory kernel. As such, here we present an algorithm that allows one to use fewer trajectories to generate the less important correlation functions while minimizing the resulting loss in accuracy for the RDM.

Any algorithm that allows for the selective convergence of the memory kernel elements should satisfy certain requirements. In particular, after specifying the RDM-level initial condition that one wishes to propagate, the algorithm should provide a prescription for apportioning a fixed total number of quantum-classical trajectories, $N_{tot}$, among the different initial conditions that need to be sampled at the the memory kernel level. Specifically, the algorithm should provide a set of normalized weights, $\left\{w\right\}$, that determine the portion of trajectories $w_m N_{tot}$ assigned to a given memory kernel level initial condition, $A_m^{\dagger}$. These weights should be chosen such that, when the time-evolved RDM is propagated from the specified initial condition, the error that arises from the statistical error in the memory kernel is minimized. In addition, the calculation of these weights should take into account two important observations. First, correlation functions with certain memory kernel-level initial conditions may contribute more to the relevant relaxation pathways than others. Second, correlation functions arising from different memory kernel-level initial conditions may require different numbers of trajectories to achieve reasonable levels of convergence. 

\begin{figure}
    \centering
    \includegraphics[width=\columnwidth]{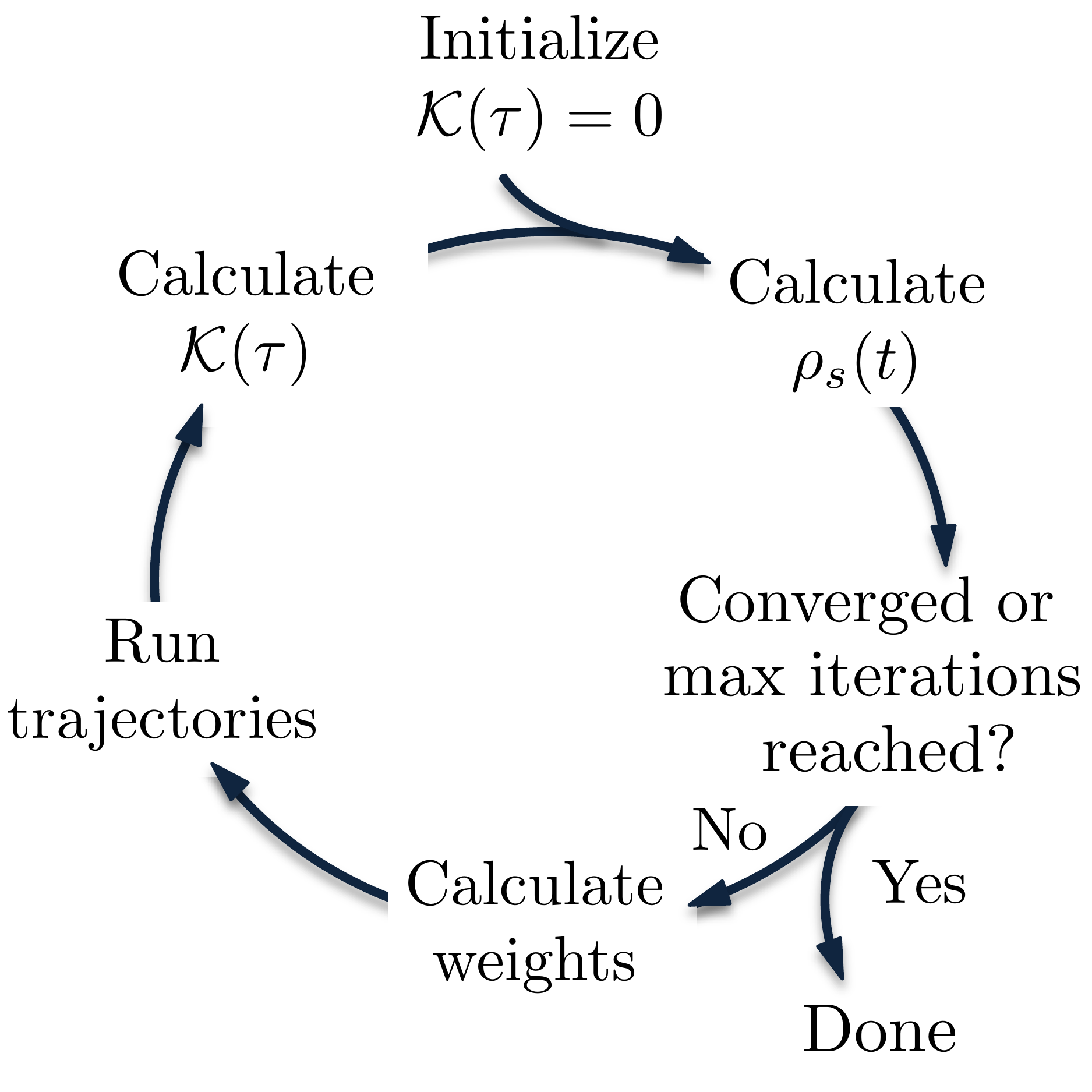}
    \caption{Overview of an iterative approach to identifying and selectively converging the memory kernel given an initial state or states of interest.} 
    \label{algorithm}
\end{figure}

To calculate the optimal weights, some knowledge of the relaxation pathway and current level of convergence of the memory kernel is crucial. Here we suggest an iterative approach, where increasingly improved estimates for the memory kernel and the RDM dynamics are used to re-compute the weights at each iteration. An overview of the algorithm is shown in Fig.~\ref{algorithm}.  We note that while here we present one particular form for the weights, the approach used here is general and does not depend on this particular choice.

To provide a form for the weights, $w_m$, we consider how a given realization of the RDM dynamics, which we call the trial RDM dynamics, deviates from its converged form, $\delta \rho_s(t) = \rho_s^{conv}(t) - \rho_s(t)$. Although in practice one does not know the converged dynamics ${\rho}_s^{conv}(t)$, considering how an unconverged dynamics may deviate from its converged form will allow us to elucidate how the convergence of the dynamics can be accelerated via selective sampling. Our goal is to reduce the absolute value of this difference from the converged dynamics for a given number of trajectories, $N_{tot}$. Because the dynamics $\rho_s(t)$, arise from integrating Eq.\ (\ref{eq:rdm_evol}), one can minimize the absolute difference between the time-derivative of the trial RDM and its converged form, $\delta \rho_s(t) = \rho_s^{conv}(t) - \rho_s(t)$. 
Because this error metric is time dependent and different for each element of the density, it is convenient to consider the mean absolute deviation, which is the absolute value of $\delta \rho_s(t)$, averaged over time and elements of the density matrix:

\begin{equation}
    \langle~ \left| \delta \rho_s \right| ~\rangle = \frac{1}{t_{max} N_s^2}\sum_{n=1}^{N_s^2} \int_0^{t_{max}} d\tau \left|~ [\delta \rho_s(\tau)]_n ~\right|,
    \label{eq:mean_absolute_deviation}
\end{equation}
where $t_{max}$ is the maximum time that is of interest of the dynamics.

Because there is no error in the initial condition, (i.e., $\delta \rho_s(0) = 0$), one can minimize the error at subsequent times by minimizing the error in the time-derivative of the RDM, $\delta\dot{\rho}_s(t) = \dot{\rho}_s^{conv}(t) - \dot{\rho_s}(t)=\frac{d}{dt} \delta \rho_s (t)$.  Again it is convenient to consider the mean absolute deviation in this quantity, which is:

\begin{equation}
    \langle~ \left| \delta \dot{\rho}_s \right| ~\rangle = \frac{1}{t_{max} N_s^2}\sum_{n=1}^{N_s^2} \int_0^{t_{max}} d\tau \left|~ [\delta \dot{\rho}_s(\tau)]_n ~\right|,
\end{equation}

An approximate upper bound of this quantity is (see Appendix \ref{App:GQME_error_analysis}),
\begin{equation}
    \abs*{\delta \dot{\rho}_s(t) } \lesssim \frac{1}{t_{max}} \sum_m \int_0^{t_{max}} dt ~ [\epsilon(t)]_m 
    \label{Eq:RDM_error}
\end{equation}
where the index $m$ runs over all initial conditions. The selective sampling error metric (see also Appendix \ref{App:GQME_error_analysis}), is approximately proportional to the error in the dynamics at time $t$ that arises from finite sampling of correlation functions with initial condition $m$. For the models considered here, this quantity is:
\begin{equation}
    [\epsilon(t)]_m = \frac{1}{\sqrt{N_m}}\int_0^t dt^{\prime}  \sum_{\beta=1}^{N_s^2} \left|\vphantom{\int_0^t}[\rho_s(t-t^{\prime})]_m \right| [\sigma_{K_1}(t^{\prime})]_{m \beta},
    \label{eq:epsilon_tilde_m}
\end{equation}
where $N_m$ is the number of trajectories that have been run in all previous iterations of the algorithm with memory kernel level initial condition $m$. The quantity $[\sigma_{K_1}(t)]_{m \beta}$ is given by Eq.\ (\ref{eq:sigma_K1}) in Appendix \ref{App:GQME_error_analysis}.

Motivated by the above considerations, the computed weights in our algorithm depend on $\rho_s(t)$, as well as the statistical error of the correlation functions that contribute to a specific elements of the memory kernel. By distributing $N_{tot}$ trajectories over $n_{step}$ steps in increments of $\Delta N = N_{tot}/n_{step}$ to different GQME-level initial conditions, we refine our estimate for these weights iteratively at each step, as described below. We repeat this procedure either until the RDM converges to a desired threshold or until $N_{tot}$ trajectories have been run. 

Using the definitions provided above, we define the un-normalized weight, $a_m$, as 
\begin{equation}
    a_m = \int_0^{t_{max}} [\epsilon(t)]_m,
    \label{eq:un_normalized_weight}
\end{equation}
The normalized weight $w_m$ is then
\begin{equation}
    w_m = \frac{a_m}{\sum_k a_k}.
    \label{eq:normalized_weight}
\end{equation}
Using these weights, we then apportion  trajectories to each memory kernel level initial condition $\hat{A}_m^{\dagger}$ in the subsequent cycle so that,
\begin{equation}
    N_m^{(j)} = w_m \times j \times \Delta N.
    \label{eq:trajectory_assignment}
\end{equation}
where $j$ is the index of the current iteration of the algorithm (for example, in the first iteration of the algorithm $j = 1$). $ N_m^{(j)}$ is therefore the number of trajectories that have been assigned to initial condition $m$ at step $j$. The algorithm can be summarized as follows: 
    \begin{enumerate}
        \item Initialize the algorithm by setting $\mathcal{K}(t) = 0$ and $ [\sigma_{\mathcal{K}_1}]_{m\beta} = 1$. 

        \item Use $\mathcal{K}(t)$ to intergrate the GQME (Eq.\ (\ref{eq:rdm_evol})) for some time range $t_{max}$, yielding $\rho_{s}(t)$ 
        
        \item Use the current guess for $\mathcal{K}(t)$ and $\rho_s(t)$ to compute the weights, $w_m$, corresponding to a given memory kernel-level initial condition $\hat{A}_m^{\dagger}$ using Eqs.\ (\ref{eq:epsilon_tilde_m})--(\ref{eq:normalized_weight}). 
        
        \item Distribute a number of MFT trajectories, $\Delta N$, among all memory kernel-level initial conditions according to Eq.\ (\ref{eq:trajectory_assignment}). 
        
        \item  Using the correlation functions $q_{3}^{(j,\pm)}(t)$ and $q_{1}^{(jk,\pm)}(t)$ obtained from all MFT trajectories that have been run so far, rebuild  $\mathcal{K}(t)$ using Eqs.\ (\ref{eq:K_K1_K3})--(\ref{eq:general_q1}). 
        
        \item  Calculate $\sigma_{\mathcal{K}_1}(t)$ using the standard deviations in the correlation functions $q_{1}^{(jk,\pm)}(t)$ (see Appendix \ref{App:GQME_error_analysis}).
        
        \item Calculate the mean absolute deviation between the previous and the current iteration of $\rho_s(t)$. If the maximum number of trajectories, $N_{tot}$ is exceeded, or if the mean absolute deviation falls below the predetermined threshold, stop. Otherwise, return to step 2.
    \end{enumerate}

A plot of the number of trajectories assigned to each initial condition for the LHCII model discussed in Section \ref{sec:results_and_discussion} is available in the SI.

\section{Simulation Details}

\subsection{Hamiltonian} \label{ssec:model_hamiltonian}
Here, we apply MF-GQME to the Frenkel exciton Hamiltonian. \cite{FMO1,Ishizaki2009b} This model Hamiltonian forms the basis of many studies of electronic energy transfer in photosynthetic light harvesting systems \cite{MayKuhn, Jang2018}. In addition to providing physical insight into these systems, the Frenkel exciton model's simplicity, i.e., its bilinear coupling to harmonic baths, allows for exact results to be calculated using, for example, the hierarchical equations of motion (HEOM) approach.\cite{Ishizaki2009} Although GQMEs, and in particular MF-GQME, are not limited to any specific form of the Hamiltonian, the availability of exact results for this model system allows us to benchmark the accuracy and efficiency of MF-GQME for multi-state problems. The Frenkel exciton Hamiltonian is of the general form given in Eq.~(\ref{eq:general-open-quantum-system-hamiltonian}), with the subsystem Hamiltonian taking the form,
\begin{equation}
    \hat{H}_s = \sum_{k=1}^{N} E^{(k)}  \ket{k}\bra{k} + \sum_{k \neq k^{\prime}} \Delta_{kk^{\prime}} \ket{k}\bra{k^{\prime}},
    \label{eq:FE_H_s}
\end{equation}
where  $E^{(k)}$ is the energy of diabatic state $\ket{k}$ and $\Delta_{kk^{\prime}}$ is the electronic coupling between states $\ket{k}$ and $\ket{k^{\prime}}$. Each diabatic state is bilinearly coupled to a bath of $N_{osc}$ independent harmonic oscillators, 
\begin{equation}\label{eq:FE_H_sb}
\begin{split}
    \hat{H}_{sb} &= \sum_{k=1}^{N_s}\sum_{j=1}^{N_{osc}} c_{j,k} \hat{R}_{j,k} \otimes \ket{k}\bra{k} \\
    &\equiv \sum_{k=1}^{N_s} \hat{\gamma}_{k} \otimes \ket{k}\bra{k},
\end{split}
\end{equation}
where $j$ indexes the bath degrees of freedom, $c_{j,k}$ is the coupling constant for a given bath oscillator, $\hat{R}_{j,k}$ is the position operator for that oscillator, and in the final equality we define $\hat{\gamma}_{k}$ to be the bath part of the system-bath coupling to state $\ket{k}$. The bath therefore consists of $N_s \times N_{osc}$ independent harmonic oscillators:
\begin{equation}
    \hat{H}_b = \frac{1}{2} \sum_{k=1}^{N_s}\sum_{j=1}^{N_{osc}}\Big[\hat{P}_{j,k}^2 +  \omega^2_{j,k}\hat{R}^2_{j,k}\Big],
    \label{eq:FE_H_b}
\end{equation}
where $\hat{P}_{j,k}$ is the momentum operator of the bath mode, and $\omega_{j,k}$ is its frequency, and the Hamiltonian is given in mass-weighted coordinates and momenta. The coupling of each diabatic state to its local bath is characterized by its spectral density,
\begin{equation}
    J_k(\omega) = \frac{\pi}{2} \sum_{j=1}^{N_{osc}} \frac{c_{j,k}^2}{\omega_{j,k}} \delta(\omega - \omega_{j,k}).
\end{equation}
For the FMO model, it is common to assume that all spectral densities are equivalent and take the Debye form,
\begin{equation}
    J(\omega) = \frac{2 \lambda \omega_c \omega}{\omega_c^2 + \omega^2},
\end{equation}
where $\lambda$ is the reorganization energy that defines the strength of the system-bath coupling and $\omega_c$ is the inverse of the characteristic timescale of the bath.  For the LHCII model, we adopt a sum of shifted Drude-Lorentz peaks \cite{kreisbeck_heom_lhcii}:
\begin{equation}
    J(\omega) = \sum_{k=1}^{N_{peaks}} \frac{\omega_{c,k}\lambda_k\omega}{\omega_{c,k}^2 + (\omega + \Omega_k)^2} + \frac{\omega_{c,k}\lambda_k\omega}{\omega_{c,k}^2 + (\omega - \Omega_k)^2}
\end{equation}
where we use the parameters of Ref.~\onlinecite{kreisbeck_heom_lhcii}: $N_{peaks} = 7$, $\omega_{c}^{-1} = \left\{ 30, 1400, 1000, 1400, 1000,
600, 1000 \right\} $ fs, $\lambda = \left\{ 130, 6, 18, 6, 16, 48, 17  \right\} \mathrm{cm^{-1}}$ , and $\Omega = \left\{240, 297, 342, 388, 518, 745, 915 \right\} \mathrm{cm^{-1}}$.

\subsection{Computational Methods}

For the FMO model, the spectral density was discretized with frequencies
\begin{equation}
    \omega_j = \omega_c \tan \left[ \frac{\pi}{2 N_{osc}} \left(j - \frac{1}{2} \right) \right].
    \label{eq:debye_discretization}
\end{equation}

For the LHCII model the spectral density was discretized by numerically solving for $\omega_j$ that satisfies
\begin{equation}
\begin{split}
    j - \frac{1}{2} = \sum_{k=1}^{N_{peaks}} \frac{N_{osc} \lambda_k}{\pi \lambda_{tot}} \Bigg[& \arctan \left( \frac{\omega_j + \Omega_k}{\omega_{c,k}} \right) \\
    &\ + \arctan \left( \frac{\omega_j - \Omega_k}{\omega_{c,k}} \right) \Bigg], \label{eq:lhcii_discretization}
\end{split}
\end{equation}
where $\lambda_{tot} = \sum_k \lambda_k$. For both Eqs. (\ref{eq:debye_discretization}) and (\ref{eq:lhcii_discretization}), $\omega_j$ is obtained by taking $j = 1, 2, \ldots, N_{osc}$. For both models, the couplings of the discrete oscillators are given by:
\begin{equation}
    c_j = \omega_j \sqrt{\frac{2\lambda_{tot}}{N_{osc}}}.
\end{equation}

For Ehrenfest trajectories, the equations of motion were solved using a split evolution method, where at each dynamical timestep, the bath is evolved through a half timestep of $\frac{\Delta t}{2}$ using mean field theory (see Appendix \ref{app:mft_dynamical_matrices}) with the velocity Verlet algorithm. \cite{Swope1982} The system was then evolved for a full timestep $\Delta t$ by numerically diagonalizing $H_s + H_{sb}(R,P)$, holding the bath positions and momenta constant. The bath was then evolved for another half timestep of $\frac{\Delta t}{2}$ using velocity Verlet, keeping the state of the electronic subsystem constant.  For the FMO model, the timestep used was $\Delta t = $0.25 fs, while for LHCII the timestep was $\Delta t = $0.05 fs.

The partial memory kernels were constructed using Eqs.~(\ref{eq:general_K3})-(\ref{eq:general_q1}), with the necessary correlation functions in Eqs.~(\ref{eq:general_q3}) and (\ref{eq:general_q1}) approximated using mean field theory as described in Appendix \ref{app:mft_dynamical_matrices}. The memory kernel was calculated at a lower time resolution (0.5 fs) than the direct MFT dynamics used to generate the underlying correlation functions, Eqs. (\ref{eq:general_q3}) and (\ref{eq:general_q1}), which was observed not to change the resulting GQME dynamics.  Additional details on the specific form these equations in the context of the Frenkel exciton model are given in Appendix \ref{app:auxiliary-kernels}. Plots of select memory kernel matrix elements that relate to transitions between the states that have significant populations for FMO and LHCII are available in the SI. 

Typically the memory kernel decays to zero after a short amount of time and hence can be truncated (set equal to zero) at all times greater than this time.  For the FMO model, this kernel cutoff time, $\tau_c$, was chosen by increasing the cutoff until the dynamics ceased to change, with $\tau_c$ between 125 and 300 fs giving very similar results. In practice, larger values of $\tau_c$ require more trajectories to converge.  For the dynamics of the FMO model shown in this work we used $\tau_c$ = 150 fs.  For the LHCII model, the dynamics remain unchanged for $\tau_c$ between 50 and 65 fs. To stably integrate the MF-GQME dynamics with cutoff times greater than 65 fs requires a large number of trajectories to achieve stable dynamics. This is because the LHCII complex is in a more strongly coupled regime, which gives rise to highly oscillatory memory kernels that require a very large number of trajectories to converge.  Such oscillatory memory kernels have been observed for the dynamics of the spin-boson model with the Debye spectral density in intermediate to strong coupling regimes \cite{Montoya2016a}. For all LHCII dynamics shown, we used $\tau_c$ = 65 fs. The instabilities observed for LHCII might be able to be alleviated by employing alternative closures,\cite{Zhang2006a,Montoya2016a,Mulvihill2019} or by using higher level quantum-classical methods to generate the memory kernel.\cite{Kelly2013, KellyMontoya2016} 

After calculating the partial memory kernels, the GQME, Eq.~(\ref{eq:rdm_evol}), was integrated using Heun's method \cite{heun_book}, with the integral in Eq.~(\ref{eq:rdm_evol}) approximated using the trapezoid rule. 

Because the sampling of bath initial conditions from which the correlation functions are generated is stochastic, there is variability in how many trajectories are required to converge the memory kernel for a given realization of uniform or selective sampling.  To evaluate our selective sampling algorithm on the LHCII model, we performed 50 separate realizations of the selective sampling algorithm, as well as 50 separate realizations of uniform sampling. All GQME dynamics shown are from the realization with the median error in the dynamics, where error in the dynamics is quantified by the mean absolute deviation in the subsystem density matrix from its converged form, Eq. (\ref{eq:mean_absolute_deviation}).  The converged form for LHCII corresponds to 65 million total trajectories, apportioned equally across all initial conditions. A plot that shows the distribution of errors of different realizations of the selective sampling procedure as a function of number of trajectories added can be found in the Supporting Information (SI). The selective sampling procedure was performed by adding a total of $\Delta N = 100$ trajectories at each iteration.

\section{Results and Discussion}
\label{sec:results_and_discussion}

Here we apply the MF-GQME method to electronic energy transfer in Frenkel exciton models of the FMO and LHCII light harvesting complexes. The FMO model consists of $N_s = 7$ subsystem states whereas $N_s = 14$ for LHCII. Therefore, for FMO there are $7^4 =2401$ time dependent matrix elements in the memory kernel, while for LHCII there are $14^{4}=38,416$. However, as described in Sec.~\ref{ssec:scaling}, all of the matrix elements can be obtained using very short simulations starting from 28 initial conditions for FMO or 105 distinct initial conditions for LHCII. 

\begin{figure}
    \centering
    \includegraphics[width=\columnwidth]{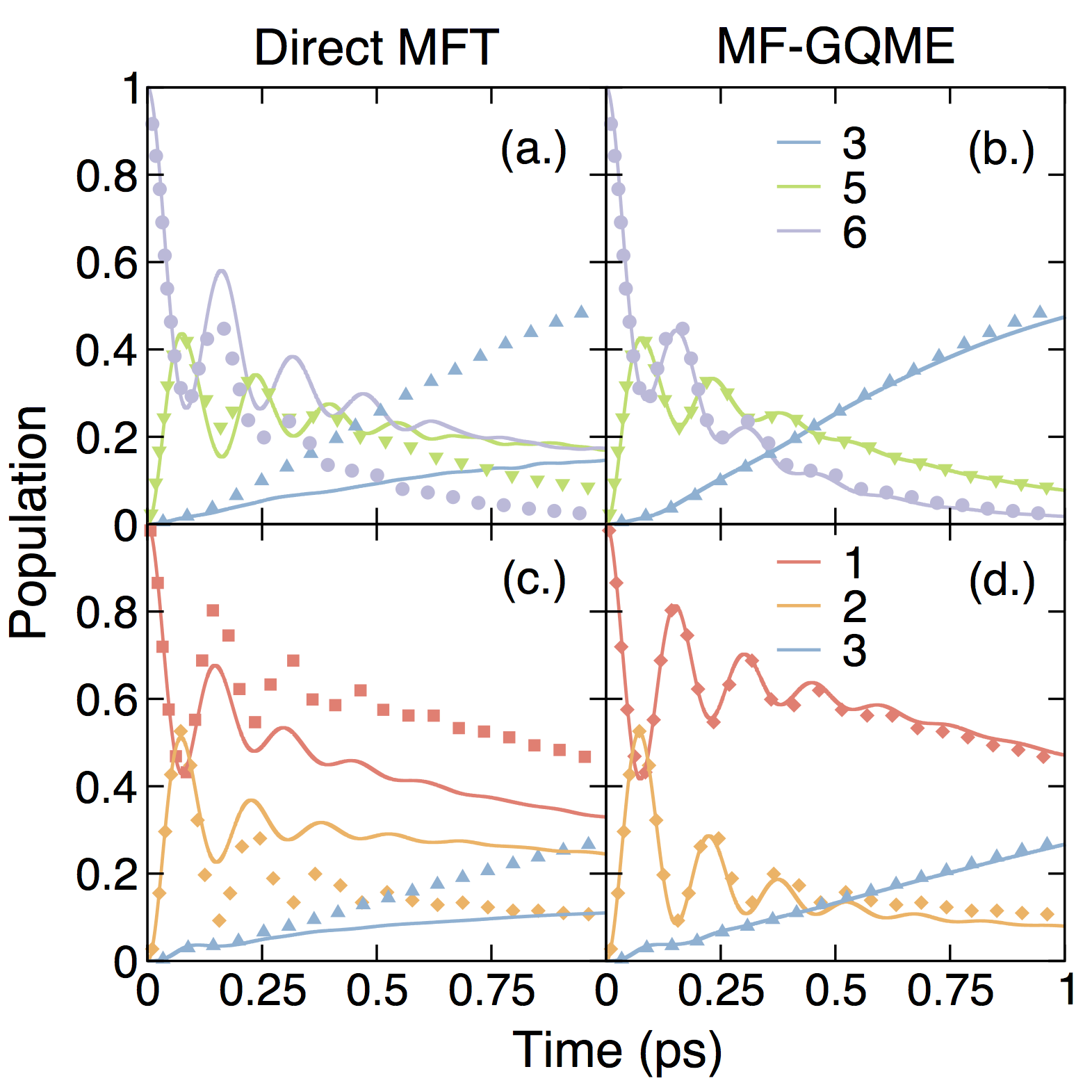}
    \caption{Populations for direct MFT and MF-GQME in the FMO model at T = 77K with 10,000 trajectories per initial condition. The model parameters are $\lambda = 35 \mathrm{cm^{-1}}$, $\omega_c^{-1} = 50 ~\mathrm{fs}$. Filled points are the exact HEOM results from Ref.~\onlinecite{Ishizaki2009b}. Panels (a.) and (b.) correspond to the initial excitation of state 6 while panels (c.) and (d.) correspond to initial excitation of state 1.}
    \label{fmo_dynamics}
\end{figure}

Figure \ref{fmo_dynamics} compares the direct MFT and MF-GQME dynamics to exact HEOM results\cite{kreisbeck_heom_lhcii} for the population relaxation following an initial excitation for the FMO model at low temperature in a regime of high nonadiabaticity ($\omega_c \gtrsim \Delta_{kk'}$ for most pairs of states). As observed in previous work for the FMO model \cite{Berkelbach2012b}, direct MFT gives qualitatively correct coherent oscillations at short times, but fails to recover the correct long-time populations. This can be explained by the high nonadiabaticity and the low temperature bath, both of which are problematic for MFT. In contrast, MF-GQME produces highly accurate results, consistent with previous observations that this scheme offers significant improvements over direct MFT in nonadiabatic regimes. \cite{Kelly2015,Pfalzgraff2015,Montoya2016a}

For this nonadiabatic problem, the memory kernel decays by a cutoff time of $\tau_c = 0.15$ ps, which is considerably shorter than the 1-2 ps lifetime of the population dynamics. However, because MF-GQME requires sampling 28 distinct initial conditions to generate the memory kernel, uniform sampling of all initial conditions in this regime, i.e., when trajectories are apportioned equally to each initial condition, requires a total of $10^5$ total trajectories ($\sim$3,500 per initial condition) to converge the memory kernel. This is in contrast to the $\sim10^4$ trajectories required to obtain a similar level of convergence with direct MFT. Hence, despite the shorter lifetime of the kernel, the larger number of trajectories required to generate the kernel means that the total cost of generating the MF-GQME dynamics shown in Fig.~\ref{fmo_dynamics} is $\sim$1.5 times greater than that of MFT. Moreover, because the memory kernel provides the dynamics for all factorizable initial conditions of the subsystem RDM, the dynamics resulting from the two distinct initial subsystem excitations shown in panels (b.) and (d.) of Fig.~\ref{fmo_dynamics} are generated using the same memory kernel. In contrast, panels (a.) and (c.) of Fig.~\ref{fmo_dynamics}, which correspond to direct MFT dynamics, require separate sets of MFT trajectories initialized from the two different subsystem initial conditions. 

\begin{figure}
    \centering
    \includegraphics[width=\columnwidth]{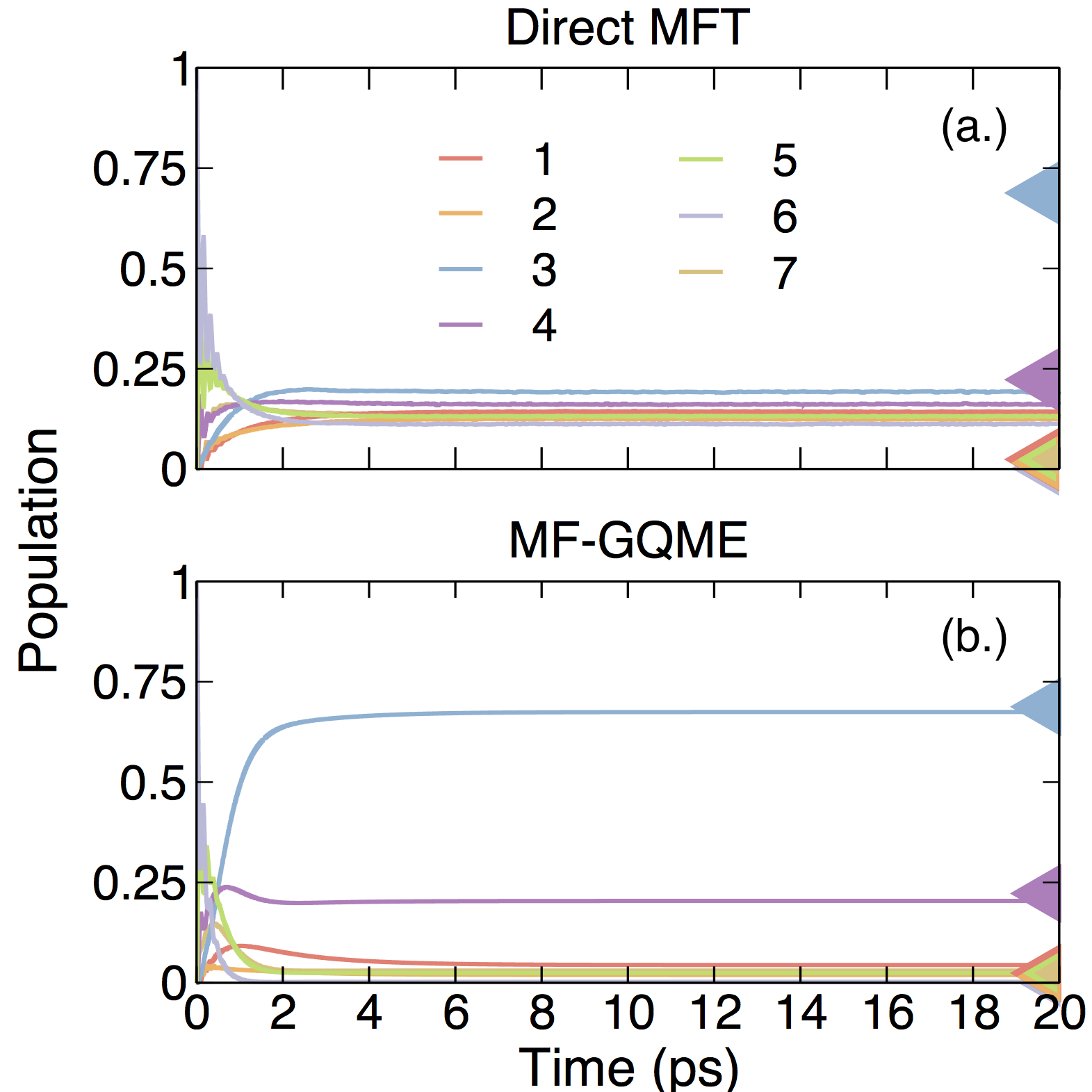}
    \caption{Long time populations in the FMO model for the parameter regime in Fig.~\ref{fmo_dynamics} with initial excitation in state 6 using direct MFT (panel (a.)) and MF-GQME (panel (b.)). The MF-GQME dynamics are generated using the same memory kernel as Fig. \ref{fmo_dynamics}. The arrows show the long-time populations that would be expected from $e^{-\beta \hat{H}_s}$.} 
    \label{long_time_population}
\end{figure}

An additional advantage of the MF-GQME scheme is that, once the memory kernel is converged, one can obtain dynamics for arbitrarily long times by integrating Eq.~(\ref{eq:rdm_evol}). In practice, the GQME propagation can be done at negligible computational cost compared to the costs of generating MFT trajectories. Figure \ref{long_time_population} shows the long time population dynamics for direct MFT and MF-GQME following initial excitation of state 6 in the same parameter regime as Fig. \ref{fmo_dynamics}. Because of the prohibitive cost of converging the exact dynamics for such long times, the arrows in Fig.~\ref{long_time_population} correspond to the Boltzmann equilibrium populations that would be expected based on the subsystem Hamiltonian, $e^{-\beta \hat{H}_s}$. As expected, due to the well-known detailed balance problems in methods that rely on a mean field description of the interaction between the electronic subsystem and the nuclei,\cite{tully_ehrenfest_balance,Tao2010,Kelly2011} MFT incorrectly predicts that all states have nearly equal population at long times, in qualitative disagreement with the expected result. In contrast, MF-GQME accurately captures the expected long-time populations. Furthermore, to converge the MFT dynamics out to the 20~ps shown in Fig.~\ref{long_time_population} requires significant additional computational effort (at least 20 times more than for 1~ps), while in the GQME approach, Eq.\ (\ref{eq:rdm_evol}) can be integrated for arbitrarily long times using the same memory kernel as Fig. \ref{fmo_dynamics}. Because integrating the GQME for longer times incurs negligible additional cost, the dynamics shown in Fig.~\ref{long_time_population} are $\sim 13$ times cheaper when generated with MF-GQME as compared with direct MFT.

A more stringent test of the MF-GQME method is the LHCII model, which has significantly more subsystem states each of which is also more strongly coupled to the bath degrees of freedom. The combination of a large number of states, stronger system-bath coupling, and the complex form of the spectral density of LHCII presents a significant challenge to exact methods. Hence, very limited benchmarks are available using the HEOM method and these required substantial computational effort and some simplification of the spectral density. \cite{kreisbeck_heom_lhcii} Here we use the form and parameterization of the LHCII Hamiltonian of Ref. \onlinecite{kreisbeck_heom_lhcii} with the J7 spectral density from that reference, and compare against the HEOM results presented therein. These HEOM results relied on the high temperature approximation,\cite{kreisbeck_heom_lhcii} and hence, while they are expected to be highly accurate, it remains unclear whether they are numerically exact. 

\begin{figure}
    \centering
    \includegraphics[width=\columnwidth]{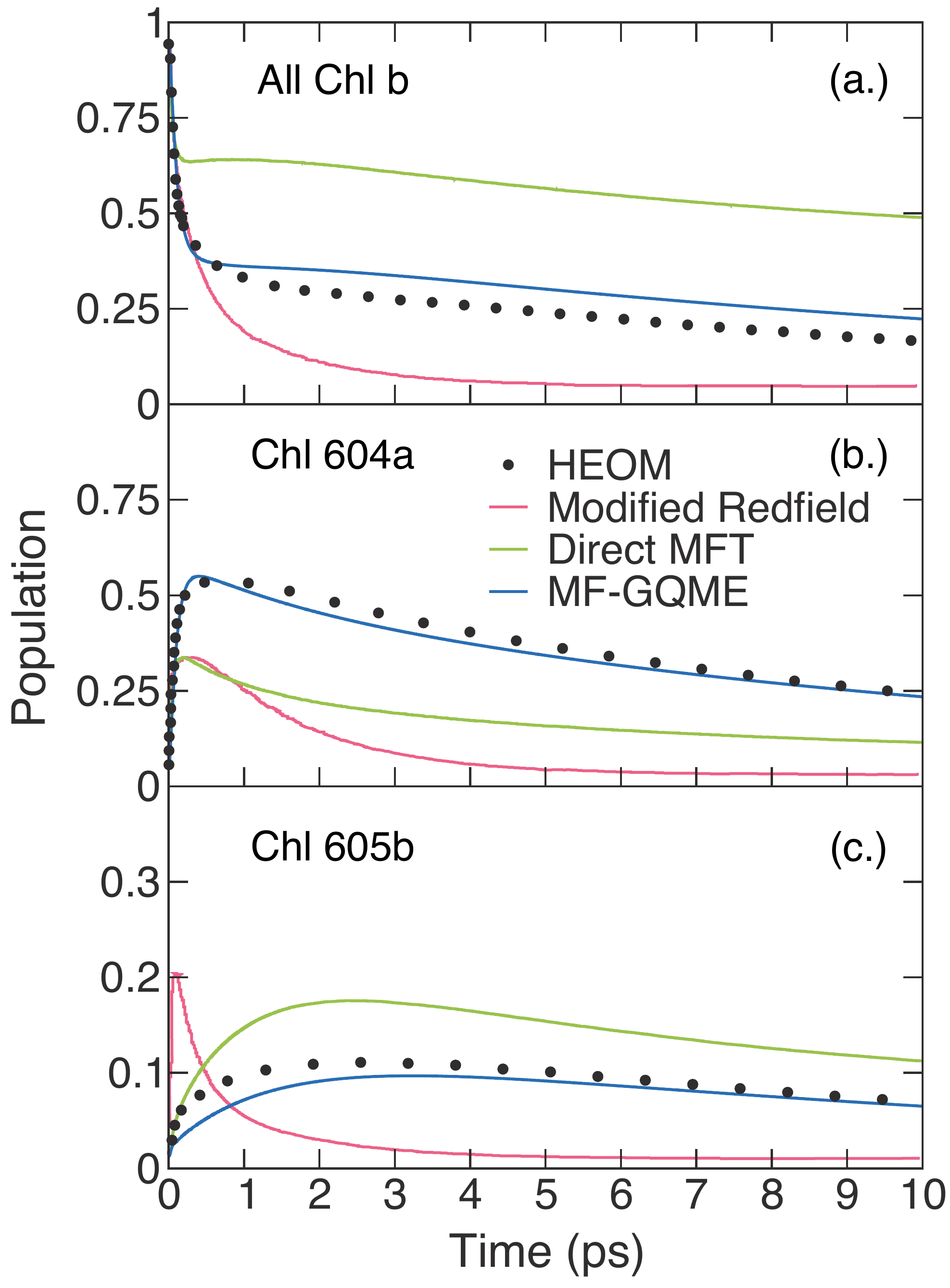}
    \caption{Population dynamics for the 14-state LHCII model of Ref. \onlinecite{kreisbeck_heom_lhcii}. Panel (a.) shows the sum of the populations of all Chl b chromophores and panels (b.) and (c.) show the populations of the Chl604a and Chl605b chromophores, respectively. The direct MFT and MF-GQME results are from this work. The direct MFT results used 100,000 trajectories and the MF-GQME results used 625,000 trajectories per initial condition ($6.5 \times 10^6$ total trajectories). For comparison, HEOM and modified Redfield results for the same model are from Ref. \onlinecite{kreisbeck_heom_lhcii}. (a.) Sum of the populations of all Chl b chromophores.  (b.) population of the Chl 604a state. (c.) Population of the Chl 605b state. The parameters of the model are those of Ref. \onlinecite{kreisbeck_heom_lhcii}}
    \label{lhcii_populations}
\end{figure}

Figure \ref{lhcii_populations} (a.) shows the dynamics for the sum of the populations of all Chl b chromophores for the LHCII model following initial excitation of the highest energy eigenstate of the subsystem Hamiltonian. For this model the exact HEOM dynamics exhibit an initial fast (sub picosecond) decay in the population of Chl b chromophores followed by a longer time component which decays over tens of picoseconds. Direct MFT fails to capture the magnitude of the short time drop in the population of Chl b chromophores, but qualitatively reproduces the longer timescale of population transfer out of this set of chromophores. As previously observed, modified Redfield dynamics, which is a major workhorse in the modelling and investigation of the excitation dynamics in photosynthetic systems, shows erroneous rapid decay within the first 1-2~ps\cite{kreisbeck_heom_lhcii}. MF-GQME, however, yields accurate short and long time behavior, in good agreement with the HEOM result. This behavior is mirrored in the populations of the Chl 604a and Chl 605b states in Fig.~\ref{lhcii_populations} (b.) and (c.) which contribute to the Chl b decay shown in (a.). Direct MFT underestimates the transfer into Chl 604a while overestimating that into Chl 605b, while modified Redfield theory leads to spuriously fast population decay of these states. In contrast, MF-GQME accurately captures the HEOM result.

\begin{figure}
    \centering
    \includegraphics[width=0.9\columnwidth]{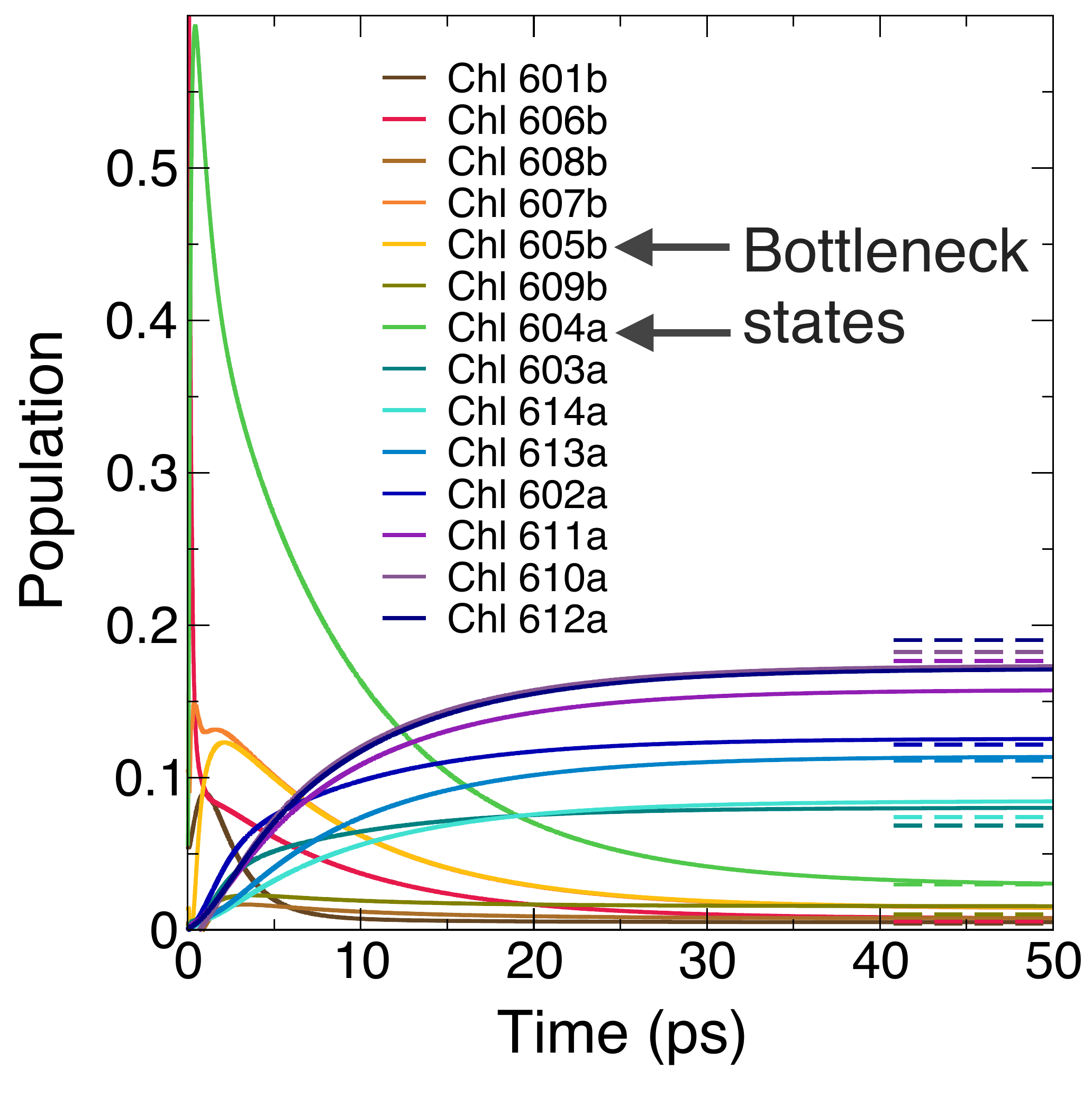}
    \caption{Long time populations in the LHCII model of ref. \onlinecite{kreisbeck_heom_lhcii} with initial excitation in the highest energy eigenstate of the subsystem Hamiltonian. The dynamics are generated using the same memory kernel as Fig. \ref{lhcii_populations}. The dashed lines show the long-time populations that would be expected from $e^{-\beta \hat{H}_s}$. }
    \label{lhcii_long_time}
\end{figure}

Figure~\ref{lhcii_long_time} shows the long-time dynamics obtained from MF-GQME using the same initial condition. In addition to capturing the exact HEOM dynamics at short times MF-GQME also obtains the intermediate timescale (picosecond) trapping in Chl 604a and Chl 605b bottleneck states, and the long timescale (tens of picoseconds) relaxation to the Chl 610a-611a-612a trimer, the Chl 613a-614a dimer, and the Chl 602a-603a dimer. \cite{Novoderezhkin2005}

Because of the stronger system-bath coupling of the LHCII model in comparison to the FMO complex, converging the memory kernel using uniform sampling of all 105 distinct initial conditions would require approximately 65 million total trajectories. Although this number of trajectories is feasible for simple models, it is intractable when more sophisticated atomistic descriptions of the environment are used. The fact that LHCII consists a large number of subsystem states makes it a prime candidate to benefit from using selective sampling to accelerate convergence. As such, we now consider the computational speedups that can be obtained by applying the selective sampling algorithm introduced in Sec.~\ref{ssec:selective} to this model. To demonstrate the utility of the selective sampling scheme, we thus compare the convergence of the GQME dynamics for a fixed total number of trajectories when using the selective sampling and the uniform sampling approaches to compute the memory kernel. See the SI for a plot showing the average number of trajectories assigned to each memory-kernel level initial condition at each step of the algorithm.

\begin{figure}
    \includegraphics[width=\columnwidth]{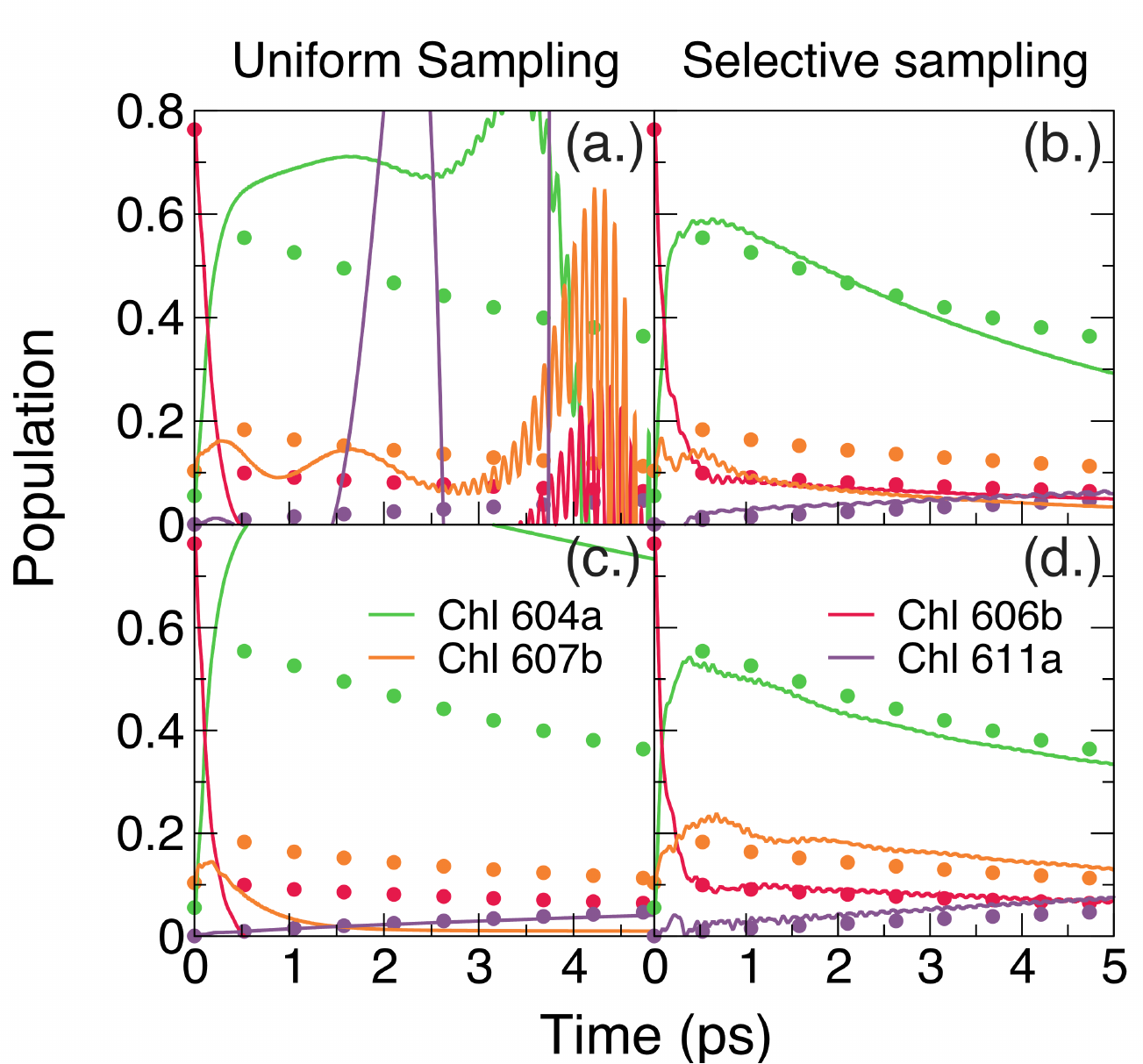}
    \caption{MF-GQME population dynamics for the LHCII model using uniform and selective sampling. Filled circles are the fully converged uniform sampling result with 65 million total trajectories. Panels (a.) and (b.) correspond to 100,000 trajectories panels (c.) and (d.) correspond to 200,000 trajectories.  The high frequency and large amplitude oscillations seen panel (a.) are caused by numerical instability due to noise in the memory kernel.}
    \label{ss_populations}
\end{figure}

Figure \ref{ss_populations} shows the populations for the median realizations of selective and uniform sampling at 100,000 and 200,000 trajectories. Already at 100,000 trajectories the selective sampling yields qualitatively correct dynamics, while the uniformly sampled memory kernel is not numerically stable, with the populations several chromophores becoming negative by $t = 1.0$ ps, and the unphysically high frequency and large amplitude oscillations seen in panel (a.) caused by numerical instability. By 200,000 trajectories, selective sampling is already in semi-quantitive agreement to the exact result, while uniform sampling produces negative populations after only $\sim$ 0.5 ps. In fact, the uniform sampling approach fully converges only after $\sim$ 65,000,000 trajectories. Hence, in this regime the selective sampling algorithm is approximately 300 times more efficient than uniform sampling.

\section{Conclusions}
In this paper, we have shown how the MF-GQME approach can be efficiently extended to treat systems where the electronic subspace consists of many states. By analyzing the symmetries of the memory kernel, we have demonstrated that, for a system consisting of $N_s$ subsystem states, the memory kernel, despite having $N_s^4$ matrix elements, can be generated using only $\mathcal{O}(N_s^2)$  quantum-classical trajectories. This scaling analysis is directly applicable to quantum-classical methods using wavefunction-based treatments of the electronic subsystem dynamics, such as MFT and FSSH. The same scaling arguments also apply to methods of generating the kernel using only the original dynamics \cite{Kidon2015a,Montoya2016a, KellyMontoya2016}, and to other methods that involve a time convolution master equation such as the transfer tensor method \cite{ttm1,ttm2}. This is in contrast to methods\cite{Sun1998, Thoss1999, Liao2002, Ananth2010, Huo2011, Hsieh2012, Ananth2013, Richardson2013} based on the Meyer-Miller-Stock-Thoss (MMST) mapping \cite{Meyer1979,Stock1997}, where the initial conditions necessary to generate all memory kernel elements could be recovered by an appropriate weighting of the variables drawn from a single Gaussian distribution of MMST variables \cite{Shi2004a,KellyMontoya2016}.

To further reduce the cost of generating the memory kernel necessary to propagate an RDM subject to specific initial conditions, we have outlined a general procedure for how one can select an optimal set of quantum-classical trajectories to converge the kernel. Employing a specific realization of a selective sampling procedure on the LHCII model that focuses resources on initial conditions that most contribute to the fast the convergence of the memory kernel, we have demonstrated that one can obtain a further $2$ orders of magnitude speed-up.  While the specific procedure used here is one possible choice, alternative ways of apportioning trajectories to specific initial conditions could lead to more efficient selective sampling schemes and represents a promising direction for future work.

By applying MF-GQME to models of excitation energy transfer in the FMO and LHCII complexes, which consist of $7$ and $14$ electronic states, respectively, we have shown that this method can provide accurate results for multi-state systems for both short-time dynamics and long-time populations at the computational cost of low-level quantum classical trajectory-based methods. Further, when memory kernels are short-lived in comparison to the electronic dynamics of interest, the MF-GQME approach usually costs less than a direct application of MFT. Moreover, once the memory kernel is generated up to the required cutoff time, $\tau_{c}$, the GQME can be propagated with linear scaling in time, making long-time dynamics easy to access.

To date, MF-GQME has previously proven successful in treating the dynamics of systems coupled to harmonic  \cite{Kelly2015, Montoya2016a, Montoya2017b} or fully atomistic \cite{Pfalzgraff2015} environments, both in\cite{Montoya2017b} and out \cite{Kelly2015,Pfalzgraff2015,Montoya2016a} of equilibrium. In the present work, we have demonstrated that this approach is also able to accurately and efficiently address the dynamics of systems where the electronic subspace consists of many states. Indeed, we have shown that MF-GQME can provide quantitatively accurate results in LHCII where commonly invoked methods, such as Redfield theory\cite{Bloch1957, Redfield1965} and its modified variant\cite{Chang1993, mukamel_redfield}, are known to fail \cite{kreisbeck_heom_lhcii}. MF-GQME thus provides a powerful tool to investigate charge and energy transfer in the condensed phase. 

\section*{Supplementary Material}
See supplementary material for plots of select memory kernel matrix elements for FMO and LHCII, a plot of the distribution of errors of different realizations of the selective sampling procedure as a function of number of trajectories added, and a plot of the number of trajectories assigned to each initial condition of the LHCII model using the selective sampling approach.

\acknowledgments
This material is based upon work supported by the U.S. Department of Energy, Office of Science, Office of Basic Energy Sciences under Award Number DE-SC0014437. T.E.M also acknowledges support from the Camille Dreyfus Teacher-Scholar Awards Program. This research used resources of the National Energy Research Scientific Computing Center, a DOE Office of Science User Facility supported by the Office of Science of the U.S. Department of Energy under Contract No. DE-AC02-05CH11231. W.C.P acknowledges support from the Melvin and Joan Lane Stanford Graduate Fellowship. We would also like to thank Stanford University and the Stanford Research Computing Center for providing computational resources and support that have contributed to these research results.

\appendix 
\section{Auxiliary kernels for a multi-level system}
\label{app:auxiliary-kernels}

In this section, we provide more detail for the derivation of the expressions for the auxiliary kernels $\mathcal{K}_1(t)$ and $\mathcal{K}_3(t)$ for the GQME that describes the dynamics of the subsystem RDM subject to a spectroscopic initial condition. These expressions are applicable to a general $N_s$ state system coupled to an arbitrary bath of the type described in Sec.~\ref{sec:theory}. We subsequently specialize our discussion to the Frenkel exciton model presented in Sec.~\ref{ssec:model_hamiltonian} given by Eqs.~(\ref{eq:FE_H_s})-(\ref{eq:FE_H_b}). 
    
We start from the general form of the Hamiltonian, Eq.~(\ref{eq:general-open-quantum-system-hamiltonian}), for an $N_s$ state electronic system coupled to a bath, which can take a harmonic (e.g., Eq.~(\ref{eq:FE_H_sb})) or atomistic form, with a general form of the system-bath coupling, given by Eq.~(\ref{eq:general_H_sb}).
    
As stated in Sec.~\ref{sec:theory}, we are interested in the nonequilibrium RDM dynamics arising from a spectroscopic or Franck-Condon initial condition, $\hat{\rho}(0) = \hat{\rho}_{s} \otimes \hat{\rho}_b^{eq}$, where the initial condition of the electronic subsystem can be expanded in terms of the Liouville basis, $\hat{\rho}_s(0) = \sum_{n} a_n \hat{A}^{\dagger}_n$, and the bath is in local equilibrium, $\hat{\rho}_b^{eq} = \frac{e^{-\beta \hat{H}_b}}{\mathrm{Tr}[e^{-\beta \hat{H}_b}]}$. To derive a particular GQME, it is also necessary to choose a projection operator. Here, we choose the $N_s$ state system generalization of the Redfield-type projection operator, shown in Ref.~\onlinecite{Montoya2016a} in the context of a two-level system. In the more conventional Nakajima-Zwanzig form of the GQME for nonequilibrium RDM dynamics \cite{Nakajima1958,Zwanzig1960a}, this projector corresponds to the commonly used Argyres-Kelley projector \cite{Argyres1964}, $\mathcal{P}$, which performs a partial trace over the bath and multiples by the canonical density of the bath, i.e., $\mathcal{P}(\hat{O}_b\hat{O}_s) = \hat{\rho}_b^{eq} \mathrm{Tr}_{b}[\hat{O}_b\hat{O}_s] = \hat{\rho}_b^{eq}\hat{O}_s \mathrm{Tr}_{b}[\hat{O}_b]$, where $\hat{O}_b$ and $\hat{O}_s$ are bath and system operators, respectively. 
    
With this form of the projection operator, and assuming that $\mathrm{Tr}[\hat{\rho}_b^{eq} \hat{\Gamma}_j] = 0$ (which can always be enforced with no loss of generality by redefining the subsystem and bath parts of the Hamiltonian: $\hat{H}_s \rightarrow  \hat{H}_s + \mathrm{Tr}[\hat{\rho}_b^{eq} H_{sb}]$ and $\hat{H}_{sb} \rightarrow  \hat{H}_{sb} - \mathrm{Tr}[\hat{\rho}_b^{eq} H_{sb}]$), it is possible to obtain the general forms for the auxiliary kernels. Focusing first on $\mathcal{K}_1(t)$,
\begin{widetext}
\begin{equation}
\begin{split}
    [\mathcal{K}_{1}(t)]_{mn} &= - \frac{1}{\hbar^2} \mathrm{Tr}\Big[(\mathcal{L}_{sb}\hat{\rho}_b^{eq}\hat{A}_m^{\dagger}) e^{i\mathcal{L}t/\hbar}(\mathcal{L}_{sb}\hat{A}_n) \Big]\\
    &= -\frac{1}{2\hbar^2} \sum_{j,k}^{N_{sb}} \Bigg\{  \mathrm{Tr}\Big[[\hat{\Gamma}_j,\hat{\rho}_b^{eq}]_{-}[\hat{S}_j,\hat{A}_m^{\dagger}]_{+} e^{i\mathcal{L}t/\hbar}[\hat{S}_{k},\hat{A}_n]_{-} \hat{\Gamma}_k \Big] + \mathrm{Tr}\Big[ [\hat{\Gamma}_j,\hat{\rho}_b^{eq}]_{+}[\hat{S}_j,\hat{A}_m^{\dagger}]_{-} e^{i\mathcal{L}t/\hbar}[\hat{S}_{k},\hat{A}_n]_{-} \hat{\Gamma}_k \Big] \Bigg\} \\
    &= \frac{1}{2\hbar^2}\sum_{j,k}^{N_{sb}} \sum_{r,s}^{N_s^2} \Bigg\{  c^{(j,+)}_{m,r} \mathrm{Tr}\Big[[\hat{\rho}_b^{eq},\hat{\Gamma}_j]_{-}\hat{A}_{r}^{\dagger} e^{i\mathcal{L}t/\hbar}\hat{A}_{s}\hat{\Gamma}_k \Big] + c_{mr}^{(j,-)}\mathrm{Tr}\Big[ [\hat{\Gamma}_j,\hat{\rho}_b^{eq}]_{+}\hat{A}_r^{\dagger} e^{i\mathcal{L}t/\hbar}\hat{A}_{s}\hat{\Gamma}_k\Big] \Bigg\}\mathbf{c}_{sn}^{(k,-)} \\
    &= \frac{1}{\hbar^2}\sum_{j,k}^{N_{sb}} \sum_{r,s}^{N_s^2} \Big\{  c_{mr}^{(j,+)}[q_1^{(jk,-)}]_{rs}(t) - c_{mr}^{(j,-)}[q_1^{(jk,+)}]_{rs}(t) \Big\} c_{sn}^{(k,-)}.
    \label{eq:general-K1-appendix}
\end{split}
\end{equation}
Using similar manipulations, one can show that, 
\begin{equation}\label{eq:general-K3-appendix}
\begin{split}
    [\mathcal{K}_{3}(t)]_{mn} &= -\frac{1}{\hbar}  \mathrm{Tr}\Big[(\mathcal{L}_{sb}\hat{\rho}_b^{eq}\hat{A}_m^{\dagger}) e^{i\mathcal{L}t/\hbar}\hat{A}_n \Big]\\
    &= \frac{1}{\hbar}\sum_{j}^{N_{sb}}\sum_{r}^{N_s^2}  \Big\{ c_{mr}^{(j,+)}[q_3^{(j,-)}]_{rn}(t)  + c_{mr}^{(j,-)}[q_3^{(j,+)}]_{rn}(t) \Big\},
\end{split}
\end{equation}
\end{widetext}
where the elements of the dynamical matrices, $q_1^{(jk,\pm)}$ and $q_3^{(j,\pm)}$, are given by Eqs.~(\ref{eq:general_q1}) and (\ref{eq:general_q3}), the elements of the static transformation matrix $c^{(j,\pm)}$ take the form,
\begin{equation}
    c^{(j,\pm)}_{mn} =  \mathrm{Tr}_s\Big[\hat{A}_{m}^{\dagger} [\hat{S}_j, \hat{A}_{n}]_{\pm}\Big].
\end{equation}
and we have used the fact that
\begin{subequations}
\begin{align}
    [\hat{S}_j, \hat{A}_n]_{\pm} &= \sum_{r}^{N_s^2} \hat{A}_{r} c_{rn}^{(j,\pm)},\\
    [\hat{S}_j, \hat{A}_{n}^{\dagger}]_{\pm} &= \pm\sum_{r}^{N_s^2}c_{nr}^{(j,\pm)} \hat{A}_r^{\dagger}.
\end{align}
\end{subequations}
Equations (\ref{eq:general-K1-appendix})-(\ref{eq:general-K3-appendix}) are general, applicable to an $N_s$ state system coupled to an arbitrary bath with system-bath coupling given by Eq.~(\ref{eq:general_H_sb}), and subject to a spectroscopic initial condition. We emphasize that such systems include electronic states coupled to harmonic or atomistic baths and system-bath interactions beyond linear coupling.

We now specialize our discussion to the Frenkel exciton Hamiltonian, Eqs.~(\ref{eq:FE_H_s})-(\ref{eq:FE_H_b}), where the bath is harmonic and can be decomposed into contributions that are local, in addition to the coupling being bilinear and local. Given the Frenkel exciton form of the Hamiltonian, the matrix elements of the static transformation matrix are given by, 
\begin{equation}
   c^{(j,\pm)}_{mn} = \delta_{j \alpha} \delta_{j \beta} \delta_{\alpha^{\prime}{\beta}^{\prime}} \pm \delta_{j \alpha^{\prime}} \delta_{j \beta^{\prime}} \delta_{\alpha \beta},
\end{equation}
where $\delta$ is the Kronecker delta.

\section{Obtaining $\mathcal{K}(t)$ from $\mathcal{K}_1(t)$ and $\mathcal{K}_3(t)$ by solving a system of linear equations}
\label{app:K_K1_K3}
Here we describe a procedure for obtaining $\mathcal{K}$ from  $\mathcal{K}_1$ and $\mathcal{K}_3$ by discretizing the self-consistent expansion and solving a system of linear equations, hence avoiding an iterative solution. The self consistency relation for the memory kernel, Eq. (\ref{eq:K_K1_K3}) can be written in discrete time as,
\begin{equation}
\begin{split}
    \mathcal{K}(t_n) \approx \mathcal{K}_1(t_n) 
    - i\frac{\Delta t}{2} \sum_{j=0}^{n-1} & \mathcal{K}_3(t_{n-j})\mathcal{K}(t_j) \\ &+ \mathcal{K}_3(t_{n-j-1})\mathcal{K}(t_{j+1}),
    \label{eq:discrete_K_decomp}
\end{split}
\end{equation}
where we have discretized the integral using the trapezoid rule. Here $t_0, t_1, \ldots, t_n$ are equally spaced discrete times, and hence $t_{j+1} - t_j \equiv \Delta t$, and $t_m - t_j = t_{m-j}$ ($m > j$).  Equation (\ref{eq:discrete_K_decomp}) can be solved to obtain an explicit expression for $\mathcal{K}(t_n)$ in terms of $\mathcal{K}_1$, $\mathcal{K}_3$, and $\mathcal{K}(t_{j})$ ($0 \leq j \leq n-1$),
\begin{widetext}
\begin{flalign*}
     \mathcal{K}(t_0) &= \mathcal{K}_1(t_0)
\end{flalign*}

\begin{flalign}
    \left[\mathbf{1} + i\frac{\Delta t}{2}\mathcal{K}_3(t_0) \right] \mathcal{K}(t_n) =  \mathcal{K}_1(t_n) - &i\frac{\Delta t}{2} \left( 
    \vphantom{\sum_{j=0}^{n-2}} 
    \mathcal{K}_3(t_1)\mathcal{K}(t_{n-1})  + \sum_{j=0}^{n-2} \mathcal{K}_3(t_{n-j})\mathcal{K}(t_j) + \mathcal{K}_3(t_{n-j-1})\mathcal{K}(t_{j+1})\right) 
    \label{eq:onestep_K1_K3}
\end{flalign}
\end{widetext}
where $n > 0$ and $\mathbf{1}$ is the identity matrix.  The above equation allows $\mathcal{K}(t_n)$ ($n > 0$) to be obtained from the $\mathcal{K}(t_{j})$ ($j < n$) using a single, non-iterative step where one solves for the unknown matrix $\mathcal{K}(t_n)$ in Eq. (\ref{eq:onestep_K1_K3}). In practice, the system of linear equations implied by Eq. (\ref{eq:onestep_K1_K3}) was solved by computing and storing the Cholesky factorization of $[\mathbf{1}  + i\frac{\Delta t}{2}\mathcal{K}_3(t_0) ]$, which can be used to solve for $\mathcal{K}(t_n)$. 

\section{MFT treatment of the dynamical matrices, $q_1^{(jk,\pm)}(t)$ and $q_3^{(j,\pm)}(t)$}
\label{app:mft_dynamical_matrices}

Here, we outline how one can calculate the dynamical matrices, $q_1^{(jk,\pm)}(t)$ and $q_3^{(j,\pm)}(t)$ in Eqs.~(\ref{eq:general_q1}) and (\ref{eq:general_q3}) using the Ehrenfest MFT method. To do this, we first review the partial transformation to Wigner phase space with respect to bath variables of a general correlation function. We then show the general MFT approximation to the dynamical matrices corresponding to a general $N_s$-state system, and specialize our discussion to the Frenkel exciton model. Finally, we provide the equations of motion for the subsystem and bath variables. 

To treat a general correlation function of a system which can be separated into a subsystem coupled to a bath, 
\begin{equation}
\begin{split}
    C_{AB}(t) &= \mathrm{Tr}[\hat{\rho} \mathcal{A} \mathcal{B}(t)]\\
    &= \mathrm{Tr}_s\mathrm{Tr}_b[\hat{\rho} \mathcal{A} \mathcal{B}(t)],
\end{split}
\end{equation}
using the Ehrenfest MFT method, we first perform a partial Wigner transformation \cite{Wigner1932, Hillery1984a} with respect to the bath variables. Here, $\hat{\rho}$ encodes the initial condition of the full system, and $\mathcal{A}$ and $\mathcal{B}$ are arbitrary operators. Upon performing the partial Wigner transformation, we can rewrite the correlation function as follows,
\begin{equation} \label{eq:correlation-function-partially-wigner-transformed}
    C_{AB}(t) = \int dX\ \mathrm{Tr}_s[(\hat{\rho} \mathcal{A})^W (\mathcal{B}(t))^W],
\end{equation}
where $X = (R,P)$ are the classical coordinates, $R$, and momenta, $P$, of the bath, and the Wigner transform of an operator, $\hat{O}$,
is defined as \cite{Imre1967, Hillery1984a}
\begin{equation} \label{eq:wigner-transform-definition}
    \hat{O}^W(X) = \int dz\ e^{\frac{iP\cdot z}{\hbar}} \langle R - \frac{z}{2} | \hat{O} | R + \frac{z}{2} \rangle.
\end{equation}
The Wigner transform of the product of two operators is \cite{Imre1967, Hillery1984a} 
\begin{equation}
    \left(\hat{A}\hat{C}\right)^{W} (X)= \hat{A}^W(X) e^{\frac{\hbar \hat{\Lambda}}{2i}}\hat{C}^{W}(X),  \label{eq:wigner_transform_op_product}
\end{equation}
where
\begin{equation}
    \hat{\Lambda} = \sum_j\left(\frac{\overleftarrow{\partial}}{\partial P_j}\frac{\overrightarrow{\partial}}{\partial R_j} - \frac{\overleftarrow{\partial}}{\partial R_j}\frac{\overrightarrow{\partial}}{\partial P_j}\right)
\end{equation}
is the Poisson bracket operator and the arrows indicate the directions in which the differential operators act. 

The MFT approximation to this correlation function amounts to replacing the time-dependent operator $(\mathcal{B}(t))^W = (e^{iHt/\hbar}\mathcal{B}e^{-iHt/\hbar})^W$ with $\mathcal{B}^W(X_t, t)$, where the classical bath degrees of freedom evolve according to the Poisson bracket subject to an additional time-dependent mean force arising from the subsystem dynamics and the quantum variables evolve according to the time-dependent Schrodinger equation subject to the time-dependent contribution of the bath. 
This allows one to rewrite the dynamical matrices, $q_1^{(jk,\pm)}(t)$ and $q_3^{(j,\pm)}(t)$, corresponding to a general $N_s$ state system in the MFT approximation as, 
\begin{subequations}
\begin{align}
    \label{eq:FE_q3_plus}
    q_{3,mn}^{(j,+)}(t) &\approx  \int dX\ \frac{([\hat{\rho}_{b}^{eq}, \hat{\Gamma}_{j}]_{+})^W(X_0)}{2}  C^{(s)}_{mn}(t), \\
    \label{eq:FE_q3_minus}
    q_{3,mn}^{(j,-)}(t) &\approx  \int dX\ \frac{([\hat{\rho}_{b}^{eq}, \hat{\Gamma}_{j}]_{-})^W(X_0)}{2}  C^{(s)}_{mn}(t), \\
    \label{eq:FE_q1_plus}
    q_{1,mn}^{(jk,+)}(t) &\approx  \int dX\ \frac{([\hat{\rho}_{b}^{eq}, \hat{\Gamma}_{j}]_{+})^W(X_0)}{2} \Gamma_{k}^{W}(X_t) C^{(s)}_{mn}(t), \\
    \label{eq:FE_q1_minus}
    q_{1,mn}^{(jk,-)}(t) &\approx \int dX\ \frac{([\hat{\rho}_{b}^{eq}, \hat{\Gamma}_{j}]_{-})^W(X_0)}{2} \Gamma_{k}^{W}(X_t) C^{(s)}_{mn}(t), 
\end{align}
\end{subequations}
where
\begin{equation}
    C^{(s)}_{mn}(t) = \mathrm{Tr}_s\left\{\hat{A}_m^{\dagger} \hat{A}_n(X_t, t)\right\}.
\end{equation}
Here we have explicitly included the dependence of operator $\hat{A}_n(t)$ on the time-dependent bath variables, $X_t$, to emphasize that the value of $C_{mn}^{(s)}(t, X_t)$ is dependent on the realization of the bath. 

For the Frenkel exciton model, where the bath is harmonic and the bath part of the system-bath coupling is linear in the bath coordinates and given by $\hat{\Gamma}_j \rightarrow \hat{\gamma}_{j} = \sum_{k} c_{j,k} \hat{R}_{j,k}$, the Wigner transform of the anticommutator and commutator of the bath canonical distribution and $\hat{\gamma}_{k}$ take simple forms,
\begin{subequations}
\begin{align}
    \frac{([\hat{\rho}_{b}^{eq}, \hat{\gamma}_{j}]_{+})^W(X_0)}{2} &= \rho_b^W(X_0) \gamma_j^W(X_0),\\ 
    \frac{([\hat{\rho}_{b}^{eq}, \hat{\gamma}_{j}]_{-})^W(X_0)}{2} &= \rho_b^W(X_0) \xi_j^W(X_0),
\end{align}
\end{subequations}
where, using Eq.~(\ref{eq:wigner_transform_op_product}), one obtains 
\begin{subequations}
\begin{align}
    \gamma_{j}^W &= \sum_{k} c_{j,k} R_{j,k},\\
    \xi_{j}^W &= - i \sum_{k} c_{j,k} P_{j,k}\frac{\tanh(\beta \hbar \omega_{j,k}/2)}{\omega_{j,k}},
\end{align}
\end{subequations}
and
\begin{equation}
\begin{split}
    \rho_b^W &= \prod_{j,k}\frac{\tanh({\beta \hbar\omega_{j,k}/2)}}{\pi} \\
    &\ \times \exp\Bigg[ - \frac{\tanh(\beta \hbar\omega_{j,k}/2)}{\hbar\omega_{j,k}}\Big[ P_{j,k}^2 + \omega_{j,k}^2 R_{j,k}^2\Big]\Bigg].
\end{split}
\end{equation}

Application of the Ehrenfest MFT approximation to the dynamics of the Frenkel exciton model leads to the evolution of the subsystem wavefunction subject to the Hamiltonian of the subsystem modulated by the time-dependent system-bath coupling,
\begin{equation}
    \frac{d}{dt} \rho_s(t, X_t) = -i\Big[ \Big(\hat{H}_s + \sum_{k} \gamma_{k}(X_t) \ket{k}\bra{k}\Big), \rho_s(t, X_t)\Big], 
\end{equation}
where the subsystem initial condition, $\rho_s(0) = \ket{\psi}\bra{\psi}$, is of pure form.

The bath, in turn, follows Hamilton's equations,
\begin{subequations}
\begin{align}
    \frac{\partial P_{j,k}}{\partial t} &= - \frac{\partial H_{b}^{Eh}}{\partial Q_{j,k}},\\
    \frac{\partial Q_{j,k}}{\partial t} &=  \frac{\partial H_{b}^{Eh}}{\partial P_{j,k}},
\end{align}
\end{subequations}
where the Ehrenfest version of the bath Hamiltonian takes the form, 
\begin{equation}
    H^{Eh}_{b}(t) = \frac{1}{2}\sum_{j,k}\Big[ P_{j,k}^2 + \omega_{j,k}^2 R_{j,k}^2 + 2c_{j,k} R_{j,k} \overline{S}_j(t) \Big],
\end{equation}
where $\overline{S}_j(t) = \mathrm{Tr}_s[\rho_s(t, X_t) \hat{S}_j]$ is the mean potential arising from the subsystem dynamics.

The correlation functions given in Eqs.~(\ref{eq:FE_q3_plus})-(\ref{eq:FE_q1_minus}) are generated using MFT and the memory kernel is then generated according to Eqs.~(\ref{eq:K_K1_K3})-(\ref{eq:general_K1}).

\section{Correlation functions with off-diagonal initial conditions using Ehrenfest mean field theory}
\label{App:off_diagonal_inicon}

In this section, we describe how the correlation functions in Eqs. (\ref{eq:general_q3}) and (\ref{eq:general_q1}) are evaluated using MFT for initial conditions that do not correspond to a single wavefunction. 

If the subsystem initial condition is diagonal, i.e. $\hat{A}_j = \ket{\alpha}\bra{\alpha^{\prime}}$ where $\alpha = \alpha^{\prime}$, then MFT is straightforward to apply because one can use the subsystem initial condition corresponding to pure state $\ket{\alpha}$, i.e. $\ket{\psi(0)} = \ket{\alpha}$. On the other hand, if $A_j$ is an off-diagonal (if $\alpha \neq \alpha^{\prime}$), the procedure is less straightforward because $A_j$ does not correspond to an initial condition that arises from a single wavefunction. Several possible approaches and subtle issues related to this are discussed in detail in the Appendices of Ref. \onlinecite{Montoya2016a}. A simple and computationally efficient approach for calculating such correlation functions involves an auxiliary wavefunction,
\begin{equation}
    \ket{\bar{\psi}} = \frac{1}{\sqrt{2}} \left( e^{i\phi}\ket{\alpha} + \ket{\alpha^{\prime}} \right)
\end{equation}
where $\phi$ is a random number uniformly sampled from the interval $[0,2\pi)$. Then consider the expression
\begin{widetext}
\begin{equation}
    \begin{aligned}
        \bar{q}(t) &= \int d \phi \int dX ~ \rho_{b,W} \Gamma_{j,W}(0)  \braket{\bar{\psi}|\alpha^{\prime}} \braket{\alpha|\bar{\psi}} \bra{\bar{\psi}} \Gamma_{k,W}(t)  A_n(t) \ket{\bar{\psi}}  \\
        &= \frac{1}{2}  \int d\phi  \int dX ~ \rho_{b,W} \Gamma_{j,W}(0)  e^{i \phi} \bra{\bar{\psi}} \Gamma_{k,W}(t)  A_n(t) \ket{\bar{\psi}} \\
        &= \frac{1}{4}  \int dX ~ \rho_{b,W} \Gamma_{j,W}(0) \bra{\alpha} \Gamma_{k,W}(t)  A_n(t) \ket{\alpha^{\prime}} \\
        &= \frac{1}{4} q(t),
    \end{aligned}
    \label{eq:off_diagonal_q}
\end{equation}
\end{widetext}
where $q(t)$ is a general memory kernel correlation function which is partially Wigner transformed over the bath degrees of freedom, such as those defined in Eqs.~(\ref{eq:FE_q3_plus})-(\ref{eq:FE_q1_minus}), and we have explicitly used that $\alpha \neq \alpha^{\prime}$. Equation (\ref{eq:off_diagonal_q}) thus provides a way to access two $q$ correlation functions using the same initial wavefunction $\ket{\bar{\psi}}$, in particular those for which $\hat{A}_m = \ket{\alpha}\bra{\alpha^{\prime}}$ and $\hat{A}_m=\ket{\alpha^{\prime}}\bra{\alpha}$. The explicit procedure for calculating such correlation functions is:
\begin{enumerate}
    \item Initialize the wavefunction $\ket{\psi(0)} = \ket{\bar{\psi}}$, sampling a new random number $\phi$ from a uniform distribution on  the interval $[0,2\pi)$. Sample the bath initial conditions from the appropriate bath distribution, $\rho_{b,W}$, as in the diagonal case.
    \item At $t=0$, evaluate and store $e^{i\phi}$ and $e^{-i\phi}$.
    \item Run each trajectory using MFT to calculate the desired observables at time $t$ in the same way as for a diagonal initial condition. Multiply the correlation function by $e^{i\phi}$ to obtain $\bar{q}_{+}(t)$ and by $e^{-i\phi}$ to obtain $\bar{q}_{-}(t)$.
    \item Repeat steps 1-4, averaging $\bar{q}_{\pm}(t)$ over an ensemble of trajectories with different bath initial conditions and different initial values of $\phi$. As shown in Eq.~(\ref{eq:off_diagonal_q}), the trajectory-averaged $\bar{q}_{+}$ corresponds $q(t)$ with $A_m = \ket{\alpha}\bra{\alpha^{\prime}}$ and $\bar{q}_{-}$ corresponds to $q(t)$ with $A_j = \ket{\alpha^{\prime}}\bra{\alpha}$.
\end{enumerate}
This approach of phase averaging over a superposition state in order to obtain the correlation function of interest is similar to that employed in the recently proposed CFBT method,\cite{cfbt} which extends Ehrenfest mean field theory.

\section{Convergence analysis of the memory kernel}
\label{App:GQME_error_analysis}

Here, we describe how to motivate our selective sampling algorithm from formal analysis of how the error in the memory kernel arising from lack of convergence translates into error in the GQME dynamics.

Our analysis starts from the GQME corresponding to the fully converged RDM, $\rho_s^{conv}(t)$, and the GQME using corresponding to a (possibly unconverged) RDM  $\rho_s(t)$. The fully converged dynamics arise from a fully converged memory kernel, $\mathcal{K}^{\mathrm{conv}}(t)$, while the unconverged dynamics arise from an unconverged memory kernel, $\mathcal{K}(t)$. A simple metric for the error in $\rho_s(t)$ is its deviation from the converged RDM,
\begin{equation}
    \delta \rho_s(t) \equiv \rho_s^{conv}(t) - \rho_s(t).   \label{eq:deviation_definition}
\end{equation}
Since this quantity can be positive or negative, it is convenient to consider its absolute value, $\left| \delta \rho_s(t) \right|$. This deviation is time dependent and is different for each element of the density. A scalar metric for the total error is the mean absolute deviation of the RDM from its converged value, with the mean taken over time and elements of the density,
\begin{equation}
    \langle \left| \delta \rho_s \right| \rangle = \frac{1}{t_{max} N_s^2}\sum_{n=1}^{N_s^2}  \left[\int_0^{t_{max}} dt \left| \delta \rho_s(t)  \right|\right]_n,
\end{equation}
where $t_{max}$ is the total amount of time over which one wants to know the RDM dynamics. Because there is no error in the initial condition, (i.e., $\delta \rho_s(0) = 0$), one can minimize the error at subsequent times by minimizing the error in the time-derivative of the RDM, 
\begin{equation}
     \begin{split}
         \delta \dot{\rho}_s(t) &\equiv \dot{\rho}_s^{\mathrm{conv}}(t) - \dot{\rho}_s(t) \\
         &= \frac{d}{dt} \left(\vphantom{\frac{d}{dt}}\delta \rho_s(t)\right).
         \label{eq:deriv_error}
     \end{split}
\end{equation}

Equation (\ref{eq:rdm_evol}) can be combined with Eq.~(\ref{eq:deriv_error}) in order to express the error in the time derivative, $ \delta \dot{\rho}_s(t)$, in terms of $\delta \rho_s(t)$ and $\delta \mathcal{K}(t)=\mathcal{K}^{\mathrm{conv}}(t) - \mathcal{K}(t)$:
\begin{equation}
     \delta \dot{\rho}_s(t) = E_1 + E_2 + E_3,
     \label{eq:error_sum}
\end{equation}
where
\begin{align}
    \label{eq:E1}
    E_1(t) &= \delta \rho_s(t) \mathcal{X},    \\
    \label{eq:E2}
    E_2(t) &= -\int_0^t d\tau \rho_s(t-\tau) \delta \mathcal{K}(\tau),\\
    \label{eq:E3}
    E_3(t) &= -\int_0^t d\tau \delta \rho_s(t-\tau) \mathcal{K}^{\mathrm{conv}}(\tau).
\end{align}

\begin{figure*}
    \centering
    \includegraphics[width=0.8\textwidth]{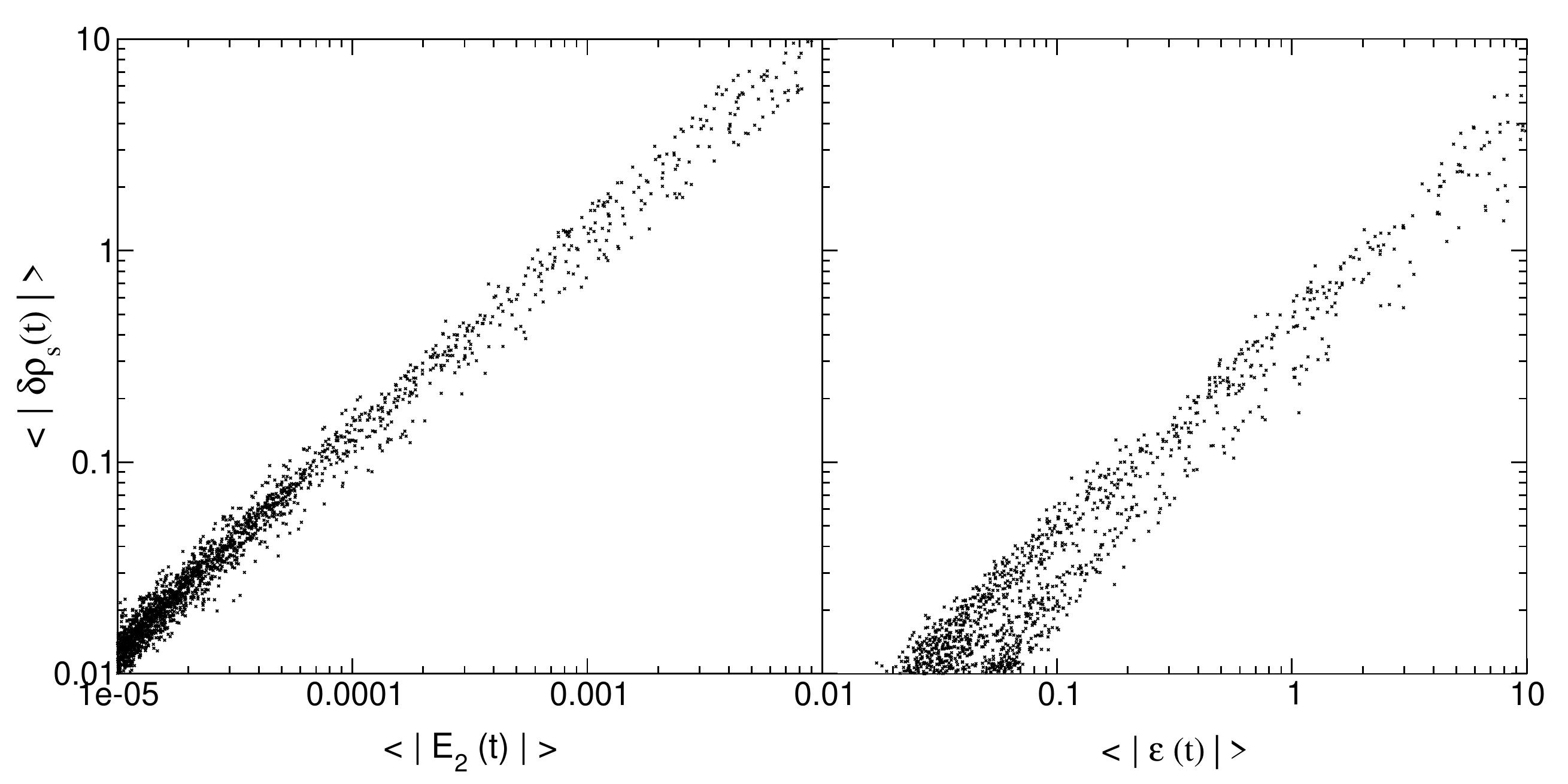}
    \caption{Correlation between the mean absolute deviation in the dynamics from its converged result $\delta \rho_s$ (left panel) and the mean absolute value of the $E_2$ and $\epsilon$ (right panel) for the LHCII model.  Each individual point corresponds to a memory kernel with a different level of convergence}
    \label{fig:error_analysis}
\end{figure*}

Of the three terms, only $E_2$ can be directly controlled by running more trajectories, since more trajectories lead to a smaller $\delta \mathcal{K}(t)$ as the correlation functions that give rise to $\mathcal{K}(t)$ become more converged. In contrast, $E_1$ and $E_3$ cannot be directly controlled because they depend on $\delta \rho_s(t)$, a quantity which arises from $\delta \mathcal{K}(t)$ and previous values of $\delta \rho_s(t)$ according to Eqs.~(\ref{eq:deriv_error})-(\ref{eq:E3}). For our choice of selective sampling algorithm we make the ansatz that the mean absolute deviation in the RDM dynamics is proportional to the mean absolute value of $E_2$, i.e., $\langle \left |\delta \rho_s \right| \rangle \propto \langle |E_2| \rangle$. Figure \ref{fig:error_analysis}a, which is generated using data from LHCII, shows strong correlation between the error in the GQME dynamics and $|E_2|$, suggests that this ansatz is valid.

In order to obtain an expression for the selective sampling weights, $w_m$, we consider how $\langle |E_2| \rangle$ is related to the error in specific elements of $\mathcal{K}(t)$,
\begin{widetext}
\begin{equation}
     \begin{split}
         \langle |E_2| \rangle &= \frac{1}{t_{max} N_s^2} \sum_{\beta=1}^{N_s^2}\int_0^{t_{max}} d\tau \left|\vphantom{\int_0^\tau} \left[ E_2(\tau)  \right]_{\beta } \right| \\
         &= \frac{1}{t_{max} N_s^2} \sum_{\alpha,\beta=1}^{N_s^2} \int_0^{t_{max}} d\tau \left|\int_0^\tau d\tau^{\prime}  [\rho_s(\tau-\tau^{\prime})]_\alpha [\delta K(\tau^{\prime})]_{\alpha \beta} \right| \\
         & \leq  \frac{1}{t_{max} N_s^2} \sum_{\alpha,\beta=1}^{N_s^2} \int_0^{t_{max}} d\tau \int_0^\tau d\tau^{\prime}  \left|\vphantom{\int_0^t}[\rho_s(\tau-\tau^{\prime})]_\alpha \right| \left|\vphantom{\int_0^t} [\delta K(\tau^{\prime})]_{\alpha \beta} \right|.
     \end{split}
\end{equation}
\end{widetext}
Using this expression to determine the error in the dynamics still requires prior knowledge of the converged kernel.  In order to use this expression to obtain an expression for the selective sampling weights, we make the additional approximation that $\left|\delta K(t)\right| \approx \left| \delta K_1(t) \right| \approx \sqrt{\frac{\sigma^2_{K_1}(t)}{N}}$ so that:
\begin{widetext}
\begin{equation}
     \langle |E_2| \rangle \lesssim \frac{1}{t_{max} N_s^2} \sum_{\alpha,\beta} \int_0^{t_{max}} d\tau \int_0^\tau d\tau^{\prime}  \left|\vphantom{\int_0^t}[\rho_s(\tau-\tau^{\prime})]_\alpha \right|  \frac{[\sigma_{K_1}(\tau^{\prime})]_{\alpha \beta}}{\sqrt{N_\alpha}},
     \label{eq:E2_inequality}
\end{equation}
\end{widetext}
where $\sigma_{\mathcal{K}_1}$ is the standard deviation in $\mathcal{K}_1$:
\begin{equation}
    \left[ \sigma_{\mathcal{K}_1 (t)} \right]_{\alpha \beta} = \sum_{\substack{j,k,\pm\\ m,n}} [c^{(j,\pm)}]_{ \alpha m} [\sigma^{(jk,\mp)}_{q_1}(t)]_{m n} [c^{(k,\pm)}]_{n \beta}.
    \label{eq:sigma_K1}
\end{equation}

Here it is useful to re-write Eq. (\ref{eq:E2_inequality}) as a sum of terms arising from different initial conditions at the memory kernel level:
\begin{equation}
    \langle |E_2| \rangle \lesssim \sum_m \int_0^{t_{max}} dt\ [\epsilon(t)]_m 
    \label{eq:decomposed_E2_inequality}
\end{equation}
where $m$ indexes the initial conditions at the memory kernel level and $[\epsilon(t)]_m$ is the sum of all the terms in Eq. (\ref{eq:E2_inequality}) arising from initial condition $m$:
\begin{widetext}
\begin{equation}
    [\epsilon(t)]_m = \frac{1}{\sqrt{N_m}}\int_0^t dt^{\prime}  \sum_{\substack{j,k,\pm\\ \alpha, \beta, n}} \left|\vphantom{\int_0^t}[\rho_s(t-t^{\prime})]_{\alpha} \right| [c^{(j,\pm)}]_{ \alpha m} [\sigma^{(jk,\mp)}_{q_1}(t)]_{m n} [c^{(k,-)}]_{n \beta},
    \label{eq:epsilon_m}
\end{equation}
\end{widetext}

Equations (\ref{eq:decomposed_E2_inequality}) and (\ref{eq:epsilon_m}) provide the basis for our choice of weights in our selective sampling algorithm.  In particular, the weight $w_m$ associated with initial condition $m$ is proportional to the time average of $[\epsilon(t) ]_m$
\begin{equation}
     w_m \propto \int_0^{t_{max}} dt\ [\epsilon(t)]_m, 
\end{equation}
This metric is an upper bound parts of the $|E_2|$ error term that arise from initial condition $m$, as shown in Eq. (\ref{eq:E2_inequality}). As shown in Fig. \ref{fig:error_analysis}b, $|E_2(t)|$ and $|\epsilon(t)|$ track the error in the GQME dynamics, $|\delta \rho_s(t)|$. Hence, a selective sampling algorithm that minimizes this term should also minimize the error in the GQME dynamics.  For the models considered here, we note that $[\epsilon(t) ]_m$ takes a somewhat simpler form because $[c^{(j,\pm)}]_{ \alpha m} = 0$ if $\alpha \neq m$ and hence

\begin{equation}
    [\epsilon(t)]_m = \frac{1}{\sqrt{N_m}}\int_0^t dt^{\prime}  \sum_{n} \left|\vphantom{\int_0^t}[\rho_s(t-t^{\prime})]_{m} \right| [\sigma_{K_1}(t)]_{m n}.
    \label{eq:FE_epsilon_m}
\end{equation}

\bibliography{mendeley,other_references}	

\begin{thebibliography}{67}%
\makeatletter
\providecommand \@ifxundefined [1]{%
 \@ifx{#1\undefined}
}%
\providecommand \@ifnum [1]{%
 \ifnum #1\expandafter \@firstoftwo
 \else \expandafter \@secondoftwo
 \fi
}%
\providecommand \@ifx [1]{%
 \ifx #1\expandafter \@firstoftwo
 \else \expandafter \@secondoftwo
 \fi
}%
\providecommand \natexlab [1]{#1}%
\providecommand \enquote  [1]{``#1''}%
\providecommand \bibnamefont  [1]{#1}%
\providecommand \bibfnamefont [1]{#1}%
\providecommand \citenamefont [1]{#1}%
\providecommand \href@noop [0]{\@secondoftwo}%
\providecommand \href [0]{\begingroup \@sanitize@url \@href}%
\providecommand \@href[1]{\@@startlink{#1}\@@href}%
\providecommand \@@href[1]{\endgroup#1\@@endlink}%
\providecommand \@sanitize@url [0]{\catcode `\\12\catcode `\$12\catcode
  `\&12\catcode `\#12\catcode `\^12\catcode `\_12\catcode `\%12\relax}%
\providecommand \@@startlink[1]{}%
\providecommand \@@endlink[0]{}%
\providecommand \url  [0]{\begingroup\@sanitize@url \@url }%
\providecommand \@url [1]{\endgroup\@href {#1}{\urlprefix }}%
\providecommand \urlprefix  [0]{URL }%
\providecommand \Eprint [0]{\href }%
\providecommand \doibase [0]{http://dx.doi.org/}%
\providecommand \selectlanguage [0]{\@gobble}%
\providecommand \bibinfo  [0]{\@secondoftwo}%
\providecommand \bibfield  [0]{\@secondoftwo}%
\providecommand \translation [1]{[#1]}%
\providecommand \BibitemOpen [0]{}%
\providecommand \bibitemStop [0]{}%
\providecommand \bibitemNoStop [0]{.\EOS\space}%
\providecommand \EOS [0]{\spacefactor3000\relax}%
\providecommand \BibitemShut  [1]{\csname bibitem#1\endcsname}%
\let\auto@bib@innerbib\@empty
\bibitem [{\citenamefont {Nakajima}(1958)}]{Nakajima1958}%
  \BibitemOpen
  \bibfield  {author} {\bibinfo {author} {\bibfnamefont {S.}~\bibnamefont
  {Nakajima}},\ }\href@noop {} {\bibfield  {journal} {\bibinfo  {journal}
  {Prog. Theor. Phys.}\ }\textbf {\bibinfo {volume} {20}},\ \bibinfo {pages}
  {948} (\bibinfo {year} {1958})}\BibitemShut {NoStop}%
\bibitem [{\citenamefont {Zwanzig}(1960)}]{Zwanzig1960a}%
  \BibitemOpen
  \bibfield  {author} {\bibinfo {author} {\bibfnamefont {R.}~\bibnamefont
  {Zwanzig}},\ }\href {\doibase 10.1063/1.1731409} {\bibfield  {journal}
  {\bibinfo  {journal} {J. Chem. Phys.}\ }\textbf {\bibinfo {volume} {33}},\
  \bibinfo {pages} {1338} (\bibinfo {year} {1960})}\BibitemShut {NoStop}%
\bibitem [{\citenamefont {Berne}\ and\ \citenamefont
  {Pecora}(1976)}]{berne_book}%
  \BibitemOpen
  \bibfield  {author} {\bibinfo {author} {\bibfnamefont {B.}~\bibnamefont
  {Berne}}\ and\ \bibinfo {author} {\bibfnamefont {R.}~\bibnamefont {Pecora}},\
  }\href@noop {} {\emph {\bibinfo {title} {Dynamic Light Scattering: With
  Applications to Chemistry, Biology, and Physics}}}\ (\bibinfo  {publisher}
  {John Wiley and Sons},\ \bibinfo {address} {New York},\ \bibinfo {year}
  {1976})\BibitemShut {NoStop}%
\bibitem [{\citenamefont {Mori}(1965)}]{Mori1965}%
  \BibitemOpen
  \bibfield  {author} {\bibinfo {author} {\bibfnamefont {H.}~\bibnamefont
  {Mori}},\ }\href {\doibase 10.1143/PTP.33.423} {\bibfield  {journal}
  {\bibinfo  {journal} {Prog. Theor. Phys.}\ }\textbf {\bibinfo {volume}
  {33}},\ \bibinfo {pages} {423} (\bibinfo {year} {1965})}\BibitemShut
  {NoStop}%
\bibitem [{\citenamefont {Bloch}(1957)}]{Bloch1957}%
  \BibitemOpen
  \bibfield  {author} {\bibinfo {author} {\bibfnamefont {F.}~\bibnamefont
  {Bloch}},\ }\href {\doibase 10.1103/PhysRev.105.1206} {\bibfield  {journal}
  {\bibinfo  {journal} {Phys. Rev.}\ }\textbf {\bibinfo {volume} {105}},\
  \bibinfo {pages} {1206} (\bibinfo {year} {1957})}\BibitemShut {NoStop}%
\bibitem [{\citenamefont {Redfield}(1965)}]{Redfield1965}%
  \BibitemOpen
  \bibfield  {author} {\bibinfo {author} {\bibfnamefont {A.~G.}\ \bibnamefont
  {Redfield}},\ }\href {\doibase 10.1016/B978-1-4832-3114-3.50007-6} {\bibfield
   {journal} {\bibinfo  {journal} {Adv. Magn. Opt. Reson.}\ }\textbf {\bibinfo
  {volume} {1}},\ \bibinfo {pages} {1} (\bibinfo {year} {1965})}\BibitemShut
  {NoStop}%
\bibitem [{\citenamefont {Chang}\ and\ \citenamefont
  {Skinner}(1993)}]{Chang1993}%
  \BibitemOpen
  \bibfield  {author} {\bibinfo {author} {\bibfnamefont {T.-M.}\ \bibnamefont
  {Chang}}\ and\ \bibinfo {author} {\bibfnamefont {J.~L.}\ \bibnamefont
  {Skinner}},\ }\href {\doibase 10.1016/0378-4371(93)90489-Q} {\bibfield
  {journal} {\bibinfo  {journal} {Physica A}\ }\textbf {\bibinfo {volume}
  {193}},\ \bibinfo {pages} {483} (\bibinfo {year} {1993})}\BibitemShut
  {NoStop}%
\bibitem [{\citenamefont {Zhang}\ \emph {et~al.}(1998)\citenamefont {Zhang},
  \citenamefont {Meier}, \citenamefont {Chernyak},\ and\ \citenamefont
  {Mukamel}}]{mukamel_redfield}%
  \BibitemOpen
  \bibfield  {author} {\bibinfo {author} {\bibfnamefont {W.~M.}\ \bibnamefont
  {Zhang}}, \bibinfo {author} {\bibfnamefont {T.}~\bibnamefont {Meier}},
  \bibinfo {author} {\bibfnamefont {V.}~\bibnamefont {Chernyak}}, \ and\
  \bibinfo {author} {\bibfnamefont {S.}~\bibnamefont {Mukamel}},\ }\href
  {\doibase 10.1063/1.476212} {\bibfield  {journal} {\bibinfo  {journal} {J.
  Chem. Phys.}\ }\textbf {\bibinfo {volume} {108}},\ \bibinfo {pages} {7763}
  (\bibinfo {year} {1998})}\BibitemShut {NoStop}%
\bibitem [{\citenamefont {Leggett}\ \emph {et~al.}(1987)\citenamefont
  {Leggett}, \citenamefont {Chakravarty}, \citenamefont {Dorsey}, \citenamefont
  {Fisher}, \citenamefont {Garg},\ and\ \citenamefont {Zwerger}}]{Leggett1987}%
  \BibitemOpen
  \bibfield  {author} {\bibinfo {author} {\bibfnamefont {A.~J.}\ \bibnamefont
  {Leggett}}, \bibinfo {author} {\bibfnamefont {S.}~\bibnamefont
  {Chakravarty}}, \bibinfo {author} {\bibfnamefont {A.~T.}\ \bibnamefont
  {Dorsey}}, \bibinfo {author} {\bibfnamefont {M.~P.~A.}\ \bibnamefont
  {Fisher}}, \bibinfo {author} {\bibfnamefont {A.}~\bibnamefont {Garg}}, \ and\
  \bibinfo {author} {\bibfnamefont {W.}~\bibnamefont {Zwerger}},\ }\href
  {\doibase 10.1103/RevModPhys.59.1} {\bibfield  {journal} {\bibinfo  {journal}
  {Rev. Mod. Phys.}\ }\textbf {\bibinfo {volume} {59}},\ \bibinfo {pages} {1}
  (\bibinfo {year} {1987})}\BibitemShut {NoStop}%
\bibitem [{\citenamefont {Dekker}(1987)}]{Dekker1987}%
  \BibitemOpen
  \bibfield  {author} {\bibinfo {author} {\bibfnamefont {H.}~\bibnamefont
  {Dekker}},\ }\href {\doibase 10.1103/PhysRevA.35.1436} {\bibfield  {journal}
  {\bibinfo  {journal} {Phys. Rev. A}\ }\textbf {\bibinfo {volume} {35}},\
  \bibinfo {pages} {1436} (\bibinfo {year} {1987})}\BibitemShut {NoStop}%
\bibitem [{\citenamefont {F\"{o}rster}(1960)}]{Forster1960}%
  \BibitemOpen
  \bibfield  {author} {\bibinfo {author} {\bibfnamefont {T.}~\bibnamefont
  {F\"{o}rster}},\ }\href@noop {} {\bibfield  {journal} {\bibinfo  {journal}
  {Radiation Research Supplement}\ }\textbf {\bibinfo {volume} {2}},\ \bibinfo
  {pages} {326} (\bibinfo {year} {1960})}\BibitemShut {NoStop}%
\bibitem [{\citenamefont {Jang}\ \emph {et~al.}(2008)\citenamefont {Jang},
  \citenamefont {Cheng}, \citenamefont {Reichman},\ and\ \citenamefont
  {Eaves}}]{Jang2008}%
  \BibitemOpen
  \bibfield  {author} {\bibinfo {author} {\bibfnamefont {S.}~\bibnamefont
  {Jang}}, \bibinfo {author} {\bibfnamefont {Y.~C.}\ \bibnamefont {Cheng}},
  \bibinfo {author} {\bibfnamefont {D.~R.}\ \bibnamefont {Reichman}}, \ and\
  \bibinfo {author} {\bibfnamefont {J.~D.}\ \bibnamefont {Eaves}},\ }\href
  {\doibase 10.1063/1.2977974} {\bibfield  {journal} {\bibinfo  {journal} {J.
  Chem. Phys.}\ }\textbf {\bibinfo {volume} {129}},\ \bibinfo {pages} {101104}
  (\bibinfo {year} {2008})}\BibitemShut {NoStop}%
\bibitem [{\citenamefont {Jang}(2011)}]{Jang2011a}%
  \BibitemOpen
  \bibfield  {author} {\bibinfo {author} {\bibfnamefont {S.}~\bibnamefont
  {Jang}},\ }\href {\doibase 10.1063/1.3608914} {\bibfield  {journal} {\bibinfo
   {journal} {J. Chem. Phys.}\ }\textbf {\bibinfo {volume} {135}},\ \bibinfo
  {pages} {34105} (\bibinfo {year} {2011})}\BibitemShut {NoStop}%
\bibitem [{\citenamefont {Nazir}(2009)}]{Nazir2009}%
  \BibitemOpen
  \bibfield  {author} {\bibinfo {author} {\bibfnamefont {A.}~\bibnamefont
  {Nazir}},\ }\href {\doibase 10.1103/PhysRevLett.103.146404} {\bibfield
  {journal} {\bibinfo  {journal} {Phys. Rev. Lett.}\ }\textbf {\bibinfo
  {volume} {103}},\ \bibinfo {pages} {146404} (\bibinfo {year}
  {2009})}\BibitemShut {NoStop}%
\bibitem [{\citenamefont {McCutcheon}\ and\ \citenamefont
  {Nazir}(2011)}]{McCutcheon2011}%
  \BibitemOpen
  \bibfield  {author} {\bibinfo {author} {\bibfnamefont {D.~P.}\ \bibnamefont
  {McCutcheon}}\ and\ \bibinfo {author} {\bibfnamefont {A.}~\bibnamefont
  {Nazir}},\ }\href {\doibase 10.1103/PhysRevB.83.165101} {\bibfield  {journal}
  {\bibinfo  {journal} {Phys. Rev. B}\ }\textbf {\bibinfo {volume} {83}},\
  \bibinfo {pages} {165101} (\bibinfo {year} {2011})}\BibitemShut {NoStop}%
\bibitem [{\citenamefont {Breuer}\ and\ \citenamefont
  {Petruccione}(2007)}]{BreuerPetruccione}%
  \BibitemOpen
  \bibfield  {author} {\bibinfo {author} {\bibfnamefont {H.-P.}\ \bibnamefont
  {Breuer}}\ and\ \bibinfo {author} {\bibfnamefont {F.}~\bibnamefont
  {Petruccione}},\ }\href@noop {} {\emph {\bibinfo {title} {{The Theory of Open
  Quantum Systems}}}}\ (\bibinfo  {publisher} {Oxford University Press},\
  \bibinfo {address} {Oxford},\ \bibinfo {year} {2007})\BibitemShut {NoStop}%
\bibitem [{\citenamefont {Nitzan}(2006)}]{NitzanBook}%
  \BibitemOpen
  \bibfield  {author} {\bibinfo {author} {\bibfnamefont {A.}~\bibnamefont
  {Nitzan}},\ }\href@noop {} {\emph {\bibinfo {title} {{Chemical Dynamics in
  Condensed Phases: Relaxation, Transfer, and Reactions in Condensed Molecular
  Systems}}}}\ (\bibinfo  {publisher} {Oxford University Press},\ \bibinfo
  {address} {New York},\ \bibinfo {year} {2006})\BibitemShut {NoStop}%
\bibitem [{\citenamefont {Shi}\ and\ \citenamefont {Geva}(2003)}]{Shi2003}%
  \BibitemOpen
  \bibfield  {author} {\bibinfo {author} {\bibfnamefont {Q.}~\bibnamefont
  {Shi}}\ and\ \bibinfo {author} {\bibfnamefont {E.}~\bibnamefont {Geva}},\
  }\href {\doibase 10.1063/1.1624830} {\bibfield  {journal} {\bibinfo
  {journal} {J. Chem. Phys.}\ }\textbf {\bibinfo {volume} {119}},\ \bibinfo
  {pages} {12063} (\bibinfo {year} {2003})}\BibitemShut {NoStop}%
\bibitem [{\citenamefont {Shi}\ and\ \citenamefont {Geva}(2004)}]{Shi2004a}%
  \BibitemOpen
  \bibfield  {author} {\bibinfo {author} {\bibfnamefont {Q.}~\bibnamefont
  {Shi}}\ and\ \bibinfo {author} {\bibfnamefont {E.}~\bibnamefont {Geva}},\
  }\href {\doibase 10.1063/1.1738109} {\bibfield  {journal} {\bibinfo
  {journal} {J. Chem. Phys.}\ }\textbf {\bibinfo {volume} {120}},\ \bibinfo
  {pages} {10647} (\bibinfo {year} {2004})}\BibitemShut {NoStop}%
\bibitem [{\citenamefont {Kelly}\ and\ \citenamefont
  {Markland}(2013)}]{Kelly2013}%
  \BibitemOpen
  \bibfield  {author} {\bibinfo {author} {\bibfnamefont {A.}~\bibnamefont
  {Kelly}}\ and\ \bibinfo {author} {\bibfnamefont {T.~E.}\ \bibnamefont
  {Markland}},\ }\href {\doibase 10.1063/1.4812355} {\bibfield  {journal}
  {\bibinfo  {journal} {J. Chem. Phys.}\ }\textbf {\bibinfo {volume} {139}},\
  \bibinfo {pages} {014104} (\bibinfo {year} {2013})}\BibitemShut {NoStop}%
\bibitem [{\citenamefont {Kelly}, \citenamefont {Brackbill},\ and\
  \citenamefont {Markland}(2015)}]{Kelly2015}%
  \BibitemOpen
  \bibfield  {author} {\bibinfo {author} {\bibfnamefont {A.}~\bibnamefont
  {Kelly}}, \bibinfo {author} {\bibfnamefont {N.}~\bibnamefont {Brackbill}}, \
  and\ \bibinfo {author} {\bibfnamefont {T.~E.}\ \bibnamefont {Markland}},\
  }\href {\doibase 10.1063/1.4913686} {\bibfield  {journal} {\bibinfo
  {journal} {J. Chem. Phys.}\ }\textbf {\bibinfo {volume} {142}},\ \bibinfo
  {pages} {094110} (\bibinfo {year} {2015})}\BibitemShut {NoStop}%
\bibitem [{\citenamefont {Pfalzgraff}, \citenamefont {Kelly},\ and\
  \citenamefont {Markland}(2015)}]{Pfalzgraff2015}%
  \BibitemOpen
  \bibfield  {author} {\bibinfo {author} {\bibfnamefont {W.~C.}\ \bibnamefont
  {Pfalzgraff}}, \bibinfo {author} {\bibfnamefont {A.}~\bibnamefont {Kelly}}, \
  and\ \bibinfo {author} {\bibfnamefont {T.~E.}\ \bibnamefont {Markland}},\
  }\href {\doibase 10.1021/acs.jpclett.5b02131} {\bibfield  {journal} {\bibinfo
   {journal} {J. Phys. Chem. Lett.}\ }\textbf {\bibinfo {volume} {6}},\
  \bibinfo {pages} {4743} (\bibinfo {year} {2015})}\BibitemShut {NoStop}%
\bibitem [{\citenamefont {Montoya-Castillo}\ and\ \citenamefont
  {Reichman}(2016)}]{Montoya2016a}%
  \BibitemOpen
  \bibfield  {author} {\bibinfo {author} {\bibfnamefont {A.}~\bibnamefont
  {Montoya-Castillo}}\ and\ \bibinfo {author} {\bibfnamefont {D.~R.}\
  \bibnamefont {Reichman}},\ }\href {\doibase 10.1063/1.4948408} {\bibfield
  {journal} {\bibinfo  {journal} {J. Chem. Phys.}\ }\textbf {\bibinfo {volume}
  {144}},\ \bibinfo {pages} {184104} (\bibinfo {year} {2016})}\BibitemShut
  {NoStop}%
\bibitem [{\citenamefont {Kelly}\ \emph {et~al.}(2016)\citenamefont {Kelly},
  \citenamefont {Montoya-Castillo}, \citenamefont {Wang},\ and\ \citenamefont
  {Markland}}]{KellyMontoya2016}%
  \BibitemOpen
  \bibfield  {author} {\bibinfo {author} {\bibfnamefont {A.}~\bibnamefont
  {Kelly}}, \bibinfo {author} {\bibfnamefont {A.}~\bibnamefont
  {Montoya-Castillo}}, \bibinfo {author} {\bibfnamefont {L.}~\bibnamefont
  {Wang}}, \ and\ \bibinfo {author} {\bibfnamefont {T.~E.}\ \bibnamefont
  {Markland}},\ }\href {\doibase 10.1063/1.4948612} {\bibfield  {journal}
  {\bibinfo  {journal} {J. Chem. Phys.}\ }\textbf {\bibinfo {volume} {144}},\
  \bibinfo {pages} {184105} (\bibinfo {year} {2016})}\BibitemShut {NoStop}%
\bibitem [{\citenamefont {Montoya-Castillo}\ and\ \citenamefont
  {Reichman}(2017)}]{Montoya2017b}%
  \BibitemOpen
  \bibfield  {author} {\bibinfo {author} {\bibfnamefont {A.}~\bibnamefont
  {Montoya-Castillo}}\ and\ \bibinfo {author} {\bibfnamefont {D.~R.}\
  \bibnamefont {Reichman}},\ }\href {\doibase 10.1063/1.4975388} {\bibfield
  {journal} {\bibinfo  {journal} {J. Chem. Phys.}\ }\textbf {\bibinfo {volume}
  {146}},\ \bibinfo {pages} {084110} (\bibinfo {year} {2017})}\BibitemShut
  {NoStop}%
\bibitem [{\citenamefont {McLachlan}(1964)}]{mft}%
  \BibitemOpen
  \bibfield  {author} {\bibinfo {author} {\bibfnamefont {A.}~\bibnamefont
  {McLachlan}},\ }\href@noop {} {\bibfield  {journal} {\bibinfo  {journal}
  {Molecular Physics}\ }\textbf {\bibinfo {volume} {8}},\ \bibinfo {pages} {39}
  (\bibinfo {year} {1964})}\BibitemShut {NoStop}%
\bibitem [{\citenamefont {Tully}(1998)}]{mft2}%
  \BibitemOpen
  \bibfield  {author} {\bibinfo {author} {\bibfnamefont {J.}~\bibnamefont
  {Tully}},\ }in\ \href@noop {} {\emph {\bibinfo {booktitle} {Classical and
  Quantum Dynamics in Condensed Phase Simulations}}},\ \bibinfo {editor}
  {edited by\ \bibinfo {editor} {\bibfnamefont {B.~J.}\ \bibnamefont {Berne}},
  \bibinfo {editor} {\bibfnamefont {G.}~\bibnamefont {Ciccotti}}, \ and\
  \bibinfo {editor} {\bibfnamefont {D.~F.}\ \bibnamefont {Coker}}}\ (\bibinfo
  {publisher} {World Scientific Publishing},\ \bibinfo {address} {Hackensack,
  NJ},\ \bibinfo {year} {1998})\ pp.\ \bibinfo {pages} {489--514}\BibitemShut
  {NoStop}%
\bibitem [{\citenamefont {Stock}(1995)}]{mft3}%
  \BibitemOpen
  \bibfield  {author} {\bibinfo {author} {\bibfnamefont {G.}~\bibnamefont
  {Stock}},\ }\href@noop {} {\bibfield  {journal} {\bibinfo  {journal} {Journal
  of Chemical Physics}\ }\textbf {\bibinfo {volume} {103}},\ \bibinfo {pages}
  {1561} (\bibinfo {year} {1995})}\BibitemShut {NoStop}%
\bibitem [{\citenamefont {Tully}\ and\ \citenamefont {Preston}(1971)}]{tully1}%
  \BibitemOpen
  \bibfield  {author} {\bibinfo {author} {\bibfnamefont {J.}~\bibnamefont
  {Tully}}\ and\ \bibinfo {author} {\bibfnamefont {R.}~\bibnamefont
  {Preston}},\ }\href@noop {} {\bibfield  {journal} {\bibinfo  {journal}
  {Journal of Chemical Physics}\ }\textbf {\bibinfo {volume} {55}},\ \bibinfo
  {pages} {562} (\bibinfo {year} {1971})}\BibitemShut {NoStop}%
\bibitem [{\citenamefont {Tully}(1990)}]{tully2}%
  \BibitemOpen
  \bibfield  {author} {\bibinfo {author} {\bibfnamefont {J.}~\bibnamefont
  {Tully}},\ }\href@noop {} {\bibfield  {journal} {\bibinfo  {journal} {Journal
  of Chemical Physics}\ }\textbf {\bibinfo {volume} {93}},\ \bibinfo {pages}
  {1061} (\bibinfo {year} {1990})}\BibitemShut {NoStop}%
\bibitem [{\citenamefont {Hammes-Schiffer}(1996)}]{Hammes-Schiffer1996}%
  \BibitemOpen
  \bibfield  {author} {\bibinfo {author} {\bibfnamefont {S.}~\bibnamefont
  {Hammes-Schiffer}},\ }\href@noop {} {\bibfield  {journal} {\bibinfo
  {journal} {J. Chem. Phys.}\ }\textbf {\bibinfo {volume} {105}},\ \bibinfo
  {pages} {2236} (\bibinfo {year} {1996})}\BibitemShut {NoStop}%
\bibitem [{\citenamefont {Adolphs}\ and\ \citenamefont {Renger}(2006)}]{FMO1}%
  \BibitemOpen
  \bibfield  {author} {\bibinfo {author} {\bibfnamefont {J.}~\bibnamefont
  {Adolphs}}\ and\ \bibinfo {author} {\bibfnamefont {T.}~\bibnamefont
  {Renger}},\ }\href@noop {} {\bibfield  {journal} {\bibinfo  {journal}
  {Biophysical Journal}\ }\textbf {\bibinfo {volume} {91}},\ \bibinfo {pages}
  {2778} (\bibinfo {year} {2006})}\BibitemShut {NoStop}%
\bibitem [{\citenamefont {Ishizaki}\ and\ \citenamefont
  {Fleming}(2009{\natexlab{a}})}]{Ishizaki2009b}%
  \BibitemOpen
  \bibfield  {author} {\bibinfo {author} {\bibfnamefont {A.}~\bibnamefont
  {Ishizaki}}\ and\ \bibinfo {author} {\bibfnamefont {G.~R.}\ \bibnamefont
  {Fleming}},\ }\href {\doibase 10.1073/pnas.0908989106} {\bibfield  {journal}
  {\bibinfo  {journal} {Proc. Natl. Acad. Sci. U.S.A.}\ }\textbf {\bibinfo
  {volume} {106}},\ \bibinfo {pages} {17255} (\bibinfo {year}
  {2009}{\natexlab{a}})}\BibitemShut {NoStop}%
\bibitem [{\citenamefont {Novoderezhkin}\ \emph {et~al.}(2005)\citenamefont
  {Novoderezhkin}, \citenamefont {Palacios}, \citenamefont {Van~Amerongen},\
  and\ \citenamefont {Van~Grondelle}}]{Novoderezhkin2005}%
  \BibitemOpen
  \bibfield  {author} {\bibinfo {author} {\bibfnamefont {V.~I.}\ \bibnamefont
  {Novoderezhkin}}, \bibinfo {author} {\bibfnamefont {M.~A.}\ \bibnamefont
  {Palacios}}, \bibinfo {author} {\bibfnamefont {H.}~\bibnamefont
  {Van~Amerongen}}, \ and\ \bibinfo {author} {\bibfnamefont {R.}~\bibnamefont
  {Van~Grondelle}},\ }\href {\doibase 10.1021/jp044082f} {\bibfield  {journal}
  {\bibinfo  {journal} {J. Phys. Chem. B}\ }\textbf {\bibinfo {volume} {109}},\
  \bibinfo {pages} {10493} (\bibinfo {year} {2005})}\BibitemShut {NoStop}%
\bibitem [{\citenamefont {van Grondelle}\ and\ \citenamefont
  {Novoderezhkin}(2006)}]{lhcii_model2}%
  \BibitemOpen
  \bibfield  {author} {\bibinfo {author} {\bibfnamefont {R.}~\bibnamefont {van
  Grondelle}}\ and\ \bibinfo {author} {\bibfnamefont {V.~I.}\ \bibnamefont
  {Novoderezhkin}},\ }\href@noop {} {\bibfield  {journal} {\bibinfo  {journal}
  {Phys. Chem. Chem. Phys.}\ }\textbf {\bibinfo {volume} {8}},\ \bibinfo
  {pages} {793} (\bibinfo {year} {2006})}\BibitemShut {NoStop}%
\bibitem [{\citenamefont {Novoderezhkin}, \citenamefont {Marin},\ and\
  \citenamefont {van Grondelle}(2011)}]{lhcii_model3}%
  \BibitemOpen
  \bibfield  {author} {\bibinfo {author} {\bibfnamefont {V.}~\bibnamefont
  {Novoderezhkin}}, \bibinfo {author} {\bibfnamefont {A.}~\bibnamefont
  {Marin}}, \ and\ \bibinfo {author} {\bibfnamefont {R.}~\bibnamefont {van
  Grondelle}},\ }\href@noop {} {\bibfield  {journal} {\bibinfo  {journal}
  {Physical Chemistry Chemical Physics}\ }\textbf {\bibinfo {volume} {13}},\
  \bibinfo {pages} {17093} (\bibinfo {year} {2011})}\BibitemShut {NoStop}%
\bibitem [{\citenamefont {Evans}\ and\ \citenamefont
  {Morriss}(2007)}]{evans_moriss}%
  \BibitemOpen
  \bibfield  {author} {\bibinfo {author} {\bibfnamefont {D.}~\bibnamefont
  {Evans}}\ and\ \bibinfo {author} {\bibfnamefont {G.}~\bibnamefont
  {Morriss}},\ }\href@noop {} {\emph {\bibinfo {title} {Statistical Mechanics
  of Nonequilibrium Liquids}}}\ (\bibinfo  {publisher} {Cambridge University
  Press},\ \bibinfo {address} {Cambridge, United Kingdom},\ \bibinfo {year}
  {2007})\BibitemShut {NoStop}%
\bibitem [{\citenamefont {May}\ and\ \citenamefont
  {K{\"{u}}hn}(2011)}]{MayKuhn}%
  \BibitemOpen
  \bibfield  {author} {\bibinfo {author} {\bibfnamefont {V.}~\bibnamefont
  {May}}\ and\ \bibinfo {author} {\bibfnamefont {O.}~\bibnamefont
  {K{\"{u}}hn}},\ }\href@noop {} {\emph {\bibinfo {title} {{Charge and Energy
  Transfer Dynamics in Molecular Systems}}}}\ (\bibinfo  {publisher}
  {Wiley-VHC},\ \bibinfo {year} {2011})\BibitemShut {NoStop}%
\bibitem [{\citenamefont {Jang}\ and\ \citenamefont
  {Mennucci}(2018)}]{Jang2018}%
  \BibitemOpen
  \bibfield  {author} {\bibinfo {author} {\bibfnamefont {S.~J.}\ \bibnamefont
  {Jang}}\ and\ \bibinfo {author} {\bibfnamefont {B.}~\bibnamefont
  {Mennucci}},\ }\href {\doibase 10.1103/RevModPhys.90.035003} {\bibfield
  {journal} {\bibinfo  {journal} {Rev. Mod. Phys.}\ }\textbf {\bibinfo {volume}
  {90}},\ \bibinfo {pages} {035003} (\bibinfo {year} {2018})}\BibitemShut
  {NoStop}%
\bibitem [{\citenamefont {Ishizaki}\ and\ \citenamefont
  {Fleming}(2009{\natexlab{b}})}]{Ishizaki2009}%
  \BibitemOpen
  \bibfield  {author} {\bibinfo {author} {\bibfnamefont {A.}~\bibnamefont
  {Ishizaki}}\ and\ \bibinfo {author} {\bibfnamefont {G.~R.}\ \bibnamefont
  {Fleming}},\ }\href {\doibase 10.1063/1.3155372} {\bibfield  {journal}
  {\bibinfo  {journal} {J. Chem. Phys.}\ }\textbf {\bibinfo {volume} {130}},\
  \bibinfo {pages} {234111} (\bibinfo {year} {2009}{\natexlab{b}})}\BibitemShut
  {NoStop}%
\bibitem [{\citenamefont {Kreisbeck}, \citenamefont {Kramer},\ and\
  \citenamefont {Aspuru-Guzik}(2014)}]{kreisbeck_heom_lhcii}%
  \BibitemOpen
  \bibfield  {author} {\bibinfo {author} {\bibfnamefont {C.}~\bibnamefont
  {Kreisbeck}}, \bibinfo {author} {\bibfnamefont {T.}~\bibnamefont {Kramer}}, \
  and\ \bibinfo {author} {\bibfnamefont {A.}~\bibnamefont {Aspuru-Guzik}},\
  }\href {\doibase 10.1021/ct500629s} {\bibfield  {journal} {\bibinfo
  {journal} {J. Chem. Theory Comput.}\ }\textbf {\bibinfo {volume} {10}},\
  \bibinfo {pages} {4045} (\bibinfo {year} {2014})}\BibitemShut {NoStop}%
\bibitem [{\citenamefont {Swope}\ \emph {et~al.}(1982)\citenamefont {Swope},
  \citenamefont {Andersen}, \citenamefont {Berens},\ and\ \citenamefont
  {Wilson}}]{Swope1982}%
  \BibitemOpen
  \bibfield  {author} {\bibinfo {author} {\bibfnamefont {W.~C.}\ \bibnamefont
  {Swope}}, \bibinfo {author} {\bibfnamefont {H.~C.}\ \bibnamefont {Andersen}},
  \bibinfo {author} {\bibfnamefont {P.~H.}\ \bibnamefont {Berens}}, \ and\
  \bibinfo {author} {\bibfnamefont {K.~R.}\ \bibnamefont {Wilson}},\ }\href
  {\doibase 10.1063/1.442716} {\bibfield  {journal} {\bibinfo  {journal} {J.
  Chem. Phys.}\ }\textbf {\bibinfo {volume} {76}},\ \bibinfo {pages} {637}
  (\bibinfo {year} {1982})}\BibitemShut {NoStop}%
\bibitem [{\citenamefont {Zhang}, \citenamefont {Ka},\ and\ \citenamefont
  {Geva}(2006)}]{Zhang2006a}%
  \BibitemOpen
  \bibfield  {author} {\bibinfo {author} {\bibfnamefont {M.-L.}\ \bibnamefont
  {Zhang}}, \bibinfo {author} {\bibfnamefont {B.~J.}\ \bibnamefont {Ka}}, \
  and\ \bibinfo {author} {\bibfnamefont {E.}~\bibnamefont {Geva}},\ }\href
  {\doibase 10.1063/1.2218342} {\bibfield  {journal} {\bibinfo  {journal} {J.
  Chem. Phys.}\ }\textbf {\bibinfo {volume} {125}},\ \bibinfo {pages} {044106}
  (\bibinfo {year} {2006})}\BibitemShut {NoStop}%
\bibitem [{\citenamefont {Mulvihill}\ \emph {et~al.}(2019)\citenamefont
  {Mulvihill}, \citenamefont {Schubert}, \citenamefont {Sun}, \citenamefont
  {Dunietz},\ and\ \citenamefont {Geva}}]{Mulvihill2019}%
  \BibitemOpen
  \bibfield  {author} {\bibinfo {author} {\bibfnamefont {E.}~\bibnamefont
  {Mulvihill}}, \bibinfo {author} {\bibfnamefont {A.}~\bibnamefont {Schubert}},
  \bibinfo {author} {\bibfnamefont {X.}~\bibnamefont {Sun}}, \bibinfo {author}
  {\bibfnamefont {B.~D.}\ \bibnamefont {Dunietz}}, \ and\ \bibinfo {author}
  {\bibfnamefont {E.}~\bibnamefont {Geva}},\ }\href@noop {} {\bibfield
  {journal} {\bibinfo  {journal} {The Journal of Chemical Physics}\ }\textbf
  {\bibinfo {volume} {150}},\ \bibinfo {pages} {034101} (\bibinfo {year}
  {2019})}\BibitemShut {NoStop}%
\bibitem [{\citenamefont {Suli}\ and\ \citenamefont
  {Mayers}(2003)}]{heun_book}%
  \BibitemOpen
  \bibfield  {author} {\bibinfo {author} {\bibfnamefont {E.}~\bibnamefont
  {Suli}}\ and\ \bibinfo {author} {\bibfnamefont {D.}~\bibnamefont {Mayers}},\
  }\href@noop {} {\emph {\bibinfo {title} {An Introduction to Numerical
  Analysis}}}\ (\bibinfo  {publisher} {Cambridge University Press},\ \bibinfo
  {address} {Cambridge, UK},\ \bibinfo {year} {2003})\BibitemShut {NoStop}%
\bibitem [{\citenamefont {Berkelbach}, \citenamefont {Markland},\ and\
  \citenamefont {Reichman}(2012)}]{Berkelbach2012b}%
  \BibitemOpen
  \bibfield  {author} {\bibinfo {author} {\bibfnamefont {T.~C.}\ \bibnamefont
  {Berkelbach}}, \bibinfo {author} {\bibfnamefont {T.~E.}\ \bibnamefont
  {Markland}}, \ and\ \bibinfo {author} {\bibfnamefont {D.~R.}\ \bibnamefont
  {Reichman}},\ }\href {\doibase 10.1063/1.3687342} {\bibfield  {journal}
  {\bibinfo  {journal} {J. Chem. Phys.}\ }\textbf {\bibinfo {volume} {136}},\
  \bibinfo {pages} {84104} (\bibinfo {year} {2012})}\BibitemShut {NoStop}%
\bibitem [{\citenamefont {Parandekar}\ and\ \citenamefont
  {Tully}(2006)}]{tully_ehrenfest_balance}%
  \BibitemOpen
  \bibfield  {author} {\bibinfo {author} {\bibfnamefont {P.}~\bibnamefont
  {Parandekar}}\ and\ \bibinfo {author} {\bibfnamefont {J.}~\bibnamefont
  {Tully}},\ }\href@noop {} {\bibfield  {journal} {\bibinfo  {journal} {J.
  Chem. Theory Comput.}\ }\textbf {\bibinfo {volume} {2}},\ \bibinfo {pages}
  {229} (\bibinfo {year} {2006})}\BibitemShut {NoStop}%
\bibitem [{\citenamefont {Tao}\ and\ \citenamefont {Miller}(2010)}]{Tao2010}%
  \BibitemOpen
  \bibfield  {author} {\bibinfo {author} {\bibfnamefont {G.}~\bibnamefont
  {Tao}}\ and\ \bibinfo {author} {\bibfnamefont {W.~H.}\ \bibnamefont
  {Miller}},\ }\href {\doibase 10.1021/jz1000825} {\bibfield  {journal}
  {\bibinfo  {journal} {J. Phys. Chem. Lett.}\ }\textbf {\bibinfo {volume}
  {1}},\ \bibinfo {pages} {891} (\bibinfo {year} {2010})}\BibitemShut {NoStop}%
\bibitem [{\citenamefont {Kelly}\ and\ \citenamefont {Rhee}(2011)}]{Kelly2011}%
  \BibitemOpen
  \bibfield  {author} {\bibinfo {author} {\bibfnamefont {A.}~\bibnamefont
  {Kelly}}\ and\ \bibinfo {author} {\bibfnamefont {Y.~M.}\ \bibnamefont
  {Rhee}},\ }\href@noop {} {\bibfield  {journal} {\bibinfo  {journal} {J. Phys.
  Chem. Lett.}\ }\textbf {\bibinfo {volume} {2}},\ \bibinfo {pages} {808}
  (\bibinfo {year} {2011})}\BibitemShut {NoStop}%
\bibitem [{\citenamefont {Kidon}, \citenamefont {Wilner},\ and\ \citenamefont
  {Rabani}(2015)}]{Kidon2015a}%
  \BibitemOpen
  \bibfield  {author} {\bibinfo {author} {\bibfnamefont {L.}~\bibnamefont
  {Kidon}}, \bibinfo {author} {\bibfnamefont {E.~Y.}\ \bibnamefont {Wilner}}, \
  and\ \bibinfo {author} {\bibfnamefont {E.}~\bibnamefont {Rabani}},\ }\href
  {\doibase 10.1063/1.4937396} {\bibfield  {journal} {\bibinfo  {journal} {J.
  Phys. Chem.}\ }\textbf {\bibinfo {volume} {143}},\ \bibinfo {pages} {234110}
  (\bibinfo {year} {2015})}\BibitemShut {NoStop}%
\bibitem [{\citenamefont {Cerrillo}\ and\ \citenamefont {Cao}(2014)}]{ttm1}%
  \BibitemOpen
  \bibfield  {author} {\bibinfo {author} {\bibfnamefont {J.}~\bibnamefont
  {Cerrillo}}\ and\ \bibinfo {author} {\bibfnamefont {J.}~\bibnamefont {Cao}},\
  }\href@noop {} {\bibfield  {journal} {\bibinfo  {journal} {Phys. Rev. Lett.}\
  }\textbf {\bibinfo {volume} {112}} (\bibinfo {year} {2014})}\BibitemShut
  {NoStop}%
\bibitem [{\citenamefont {Gelzinis}, \citenamefont {Rybakovas},\ and\
  \citenamefont {Valkunas}(2017)}]{ttm2}%
  \BibitemOpen
  \bibfield  {author} {\bibinfo {author} {\bibfnamefont {A.}~\bibnamefont
  {Gelzinis}}, \bibinfo {author} {\bibfnamefont {E.}~\bibnamefont {Rybakovas}},
  \ and\ \bibinfo {author} {\bibfnamefont {L.}~\bibnamefont {Valkunas}},\
  }\href@noop {} {\bibfield  {journal} {\bibinfo  {journal} {J. Chem. Phys.}\
  }\textbf {\bibinfo {volume} {147}},\ \bibinfo {pages} {234108} (\bibinfo
  {year} {2017})}\BibitemShut {NoStop}%
\bibitem [{\citenamefont {Sun}, \citenamefont {Wang},\ and\ \citenamefont
  {Miller}(1998)}]{Sun1998}%
  \BibitemOpen
  \bibfield  {author} {\bibinfo {author} {\bibfnamefont {X.}~\bibnamefont
  {Sun}}, \bibinfo {author} {\bibfnamefont {H.}~\bibnamefont {Wang}}, \ and\
  \bibinfo {author} {\bibfnamefont {W.~H.}\ \bibnamefont {Miller}},\ }\href
  {\doibase 10.1063/1.477389} {\bibfield  {journal} {\bibinfo  {journal} {J.
  Chem. Phys.}\ }\textbf {\bibinfo {volume} {109}},\ \bibinfo {pages} {7064}
  (\bibinfo {year} {1998})}\BibitemShut {NoStop}%
\bibitem [{\citenamefont {Thoss}\ and\ \citenamefont
  {Stock}(1999)}]{Thoss1999}%
  \BibitemOpen
  \bibfield  {author} {\bibinfo {author} {\bibfnamefont {M.}~\bibnamefont
  {Thoss}}\ and\ \bibinfo {author} {\bibfnamefont {G.}~\bibnamefont {Stock}},\
  }\href {\doibase 10.1103/PhysRevA.59.64} {\bibfield  {journal} {\bibinfo
  {journal} {Phys. Rev. A}\ }\textbf {\bibinfo {volume} {59}},\ \bibinfo
  {pages} {64} (\bibinfo {year} {1999})}\BibitemShut {NoStop}%
\bibitem [{\citenamefont {Liao}\ and\ \citenamefont {Voth}(2002)}]{Liao2002}%
  \BibitemOpen
  \bibfield  {author} {\bibinfo {author} {\bibfnamefont {J.-L.}\ \bibnamefont
  {Liao}}\ and\ \bibinfo {author} {\bibfnamefont {G.~A.}\ \bibnamefont
  {Voth}},\ }\href {\doibase 10.1021/jp020978d} {\bibfield  {journal} {\bibinfo
   {journal} {J. Phys. Chem. B}\ }\textbf {\bibinfo {volume} {106}},\ \bibinfo
  {pages} {8449} (\bibinfo {year} {2002})}\BibitemShut {NoStop}%
\bibitem [{\citenamefont {Ananth}\ and\ \citenamefont
  {Miller}(2010)}]{Ananth2010}%
  \BibitemOpen
  \bibfield  {author} {\bibinfo {author} {\bibfnamefont {N.}~\bibnamefont
  {Ananth}}\ and\ \bibinfo {author} {\bibfnamefont {T.~F.}\ \bibnamefont
  {Miller}},\ }\href {\doibase 10.1063/1.3511700} {\bibfield  {journal}
  {\bibinfo  {journal} {J. Chem. Phys.}\ }\textbf {\bibinfo {volume} {133}},\
  \bibinfo {pages} {234103} (\bibinfo {year} {2010})}\BibitemShut {NoStop}%
\bibitem [{\citenamefont {Huo}\ and\ \citenamefont {Coker}(2011)}]{Huo2011}%
  \BibitemOpen
  \bibfield  {author} {\bibinfo {author} {\bibfnamefont {P.}~\bibnamefont
  {Huo}}\ and\ \bibinfo {author} {\bibfnamefont {D.~F.}\ \bibnamefont
  {Coker}},\ }\href {\doibase 10.1063/1.3664763} {\bibfield  {journal}
  {\bibinfo  {journal} {J. Chem. Phys.}\ }\textbf {\bibinfo {volume} {135}},\
  \bibinfo {pages} {201101} (\bibinfo {year} {2011})}\BibitemShut {NoStop}%
\bibitem [{\citenamefont {Hsieh}\ and\ \citenamefont
  {Kapral}(2012)}]{Hsieh2012}%
  \BibitemOpen
  \bibfield  {author} {\bibinfo {author} {\bibfnamefont {C.~Y.}\ \bibnamefont
  {Hsieh}}\ and\ \bibinfo {author} {\bibfnamefont {R.}~\bibnamefont {Kapral}},\
  }\href {\doibase 10.1063/1.4736841} {\bibfield  {journal} {\bibinfo
  {journal} {J. Chem. Phys.}\ }\textbf {\bibinfo {volume} {137}},\ \bibinfo
  {pages} {22A507} (\bibinfo {year} {2012})}\BibitemShut {NoStop}%
\bibitem [{\citenamefont {Ananth}(2013)}]{Ananth2013}%
  \BibitemOpen
  \bibfield  {author} {\bibinfo {author} {\bibfnamefont {N.}~\bibnamefont
  {Ananth}},\ }\href {\doibase 10.1063/1.4821590} {\bibfield  {journal}
  {\bibinfo  {journal} {J. Chem. Phys.}\ }\textbf {\bibinfo {volume} {139}},\
  \bibinfo {pages} {124102} (\bibinfo {year} {2013})}\BibitemShut {NoStop}%
\bibitem [{\citenamefont {Richardson}\ and\ \citenamefont
  {Thoss}(2013)}]{Richardson2013}%
  \BibitemOpen
  \bibfield  {author} {\bibinfo {author} {\bibfnamefont {J.~O.}\ \bibnamefont
  {Richardson}}\ and\ \bibinfo {author} {\bibfnamefont {M.}~\bibnamefont
  {Thoss}},\ }\href {\doibase 10.1063/1.4816124} {\bibfield  {journal}
  {\bibinfo  {journal} {J. Chem. Phys.}\ }\textbf {\bibinfo {volume} {139}},\
  \bibinfo {pages} {31102} (\bibinfo {year} {2013})}\BibitemShut {NoStop}%
\bibitem [{\citenamefont {Meyer}\ and\ \citenamefont
  {Miller}(1979)}]{Meyer1979}%
  \BibitemOpen
  \bibfield  {author} {\bibinfo {author} {\bibfnamefont {H.-D.}\ \bibnamefont
  {Meyer}}\ and\ \bibinfo {author} {\bibfnamefont {W.~H.}\ \bibnamefont
  {Miller}},\ }\href {\doibase 10.1063/1.437910} {\bibfield  {journal}
  {\bibinfo  {journal} {J. Chem. Phys.}\ }\textbf {\bibinfo {volume} {70}},\
  \bibinfo {pages} {3214} (\bibinfo {year} {1979})}\BibitemShut {NoStop}%
\bibitem [{\citenamefont {Stock}\ and\ \citenamefont
  {Thoss}(1997)}]{Stock1997}%
  \BibitemOpen
  \bibfield  {author} {\bibinfo {author} {\bibfnamefont {G.}~\bibnamefont
  {Stock}}\ and\ \bibinfo {author} {\bibfnamefont {M.}~\bibnamefont {Thoss}},\
  }\href {\doibase 10.1103/PhysRevLett.78.578} {\bibfield  {journal} {\bibinfo
  {journal} {Phys. Rev. Lett.}\ }\textbf {\bibinfo {volume} {78}},\ \bibinfo
  {pages} {578} (\bibinfo {year} {1997})}\BibitemShut {NoStop}%
\bibitem [{\citenamefont {Argyres}\ and\ \citenamefont
  {Kelley}(1964)}]{Argyres1964}%
  \BibitemOpen
  \bibfield  {author} {\bibinfo {author} {\bibfnamefont {P.~N.}\ \bibnamefont
  {Argyres}}\ and\ \bibinfo {author} {\bibfnamefont {P.~L.}\ \bibnamefont
  {Kelley}},\ }\href {\doibase 10.1103/PhysRev.134.A98} {\bibfield  {journal}
  {\bibinfo  {journal} {Phys. Rev.}\ }\textbf {\bibinfo {volume} {134}},\
  \bibinfo {pages} {A98} (\bibinfo {year} {1964})}\BibitemShut {NoStop}%
\bibitem [{\citenamefont {Wigner}(1932)}]{Wigner1932}%
  \BibitemOpen
  \bibfield  {author} {\bibinfo {author} {\bibfnamefont {E.}~\bibnamefont
  {Wigner}},\ }\href {\doibase 10.1017/CBO9781107415324.004} {\bibfield
  {journal} {\bibinfo  {journal} {Phys. Rev.}\ }\textbf {\bibinfo {volume}
  {40}},\ \bibinfo {pages} {749} (\bibinfo {year} {1932})}\BibitemShut
  {NoStop}%
\bibitem [{\citenamefont {Hillery}\ \emph {et~al.}(1984)\citenamefont
  {Hillery}, \citenamefont {O'Connell}, \citenamefont {Scully},\ and\
  \citenamefont {Wigner}}]{Hillery1984a}%
  \BibitemOpen
  \bibfield  {author} {\bibinfo {author} {\bibfnamefont {M.}~\bibnamefont
  {Hillery}}, \bibinfo {author} {\bibfnamefont {R.~F.}\ \bibnamefont
  {O'Connell}}, \bibinfo {author} {\bibfnamefont {M.~O.}\ \bibnamefont
  {Scully}}, \ and\ \bibinfo {author} {\bibfnamefont {E.~P.}\ \bibnamefont
  {Wigner}},\ }\href {\doibase 10.1016/0370-1573(84)90160-1} {\bibfield
  {journal} {\bibinfo  {journal} {Phys. Rep.}\ }\textbf {\bibinfo {volume}
  {106}},\ \bibinfo {pages} {121} (\bibinfo {year} {1984})}\BibitemShut
  {NoStop}%
\bibitem [{\citenamefont {Imre}(1967)}]{Imre1967}%
  \BibitemOpen
  \bibfield  {author} {\bibinfo {author} {\bibfnamefont {K.}~\bibnamefont
  {Imre}},\ }\href {\doibase 10.1063/1.1705323} {\bibfield  {journal} {\bibinfo
   {journal} {J. Math. Phys.}\ }\textbf {\bibinfo {volume} {8}},\ \bibinfo
  {pages} {1097} (\bibinfo {year} {1967})}\BibitemShut {NoStop}%
\bibitem [{\citenamefont {Sato}, \citenamefont {Kelly},\ and\ \citenamefont
  {Rubio}(2018)}]{cfbt}%
  \BibitemOpen
  \bibfield  {author} {\bibinfo {author} {\bibfnamefont {S.~A.}\ \bibnamefont
  {Sato}}, \bibinfo {author} {\bibfnamefont {A.}~\bibnamefont {Kelly}}, \ and\
  \bibinfo {author} {\bibfnamefont {A.}~\bibnamefont {Rubio}},\ }\href@noop {}
  {\bibfield  {journal} {\bibinfo  {journal} {Phys. Rev. B}\ }\textbf {\bibinfo
  {volume} {97}} (\bibinfo {year} {2018})}\BibitemShut {NoStop}%
\end{thebibliography}%

\end{document}